\documentclass[preprint2,twoside]{hwo}

\usepackage{xcolor}
\usepackage{graphicx}

\input{hwo.h}

\begin{document}

\title{\textbf{\LARGE Detecting alien living worlds and photosynthetic life using imaging polarimetry with the HWO coronagraph
}}

\author {\textbf{\large 
Svetlana Berdyugina$^{1}$, 
Lucas Patty$^{2}$, 
Jonathan Grone$^{2}$,
Brice Demory$^{2}$,
Kim Bott$^{3}$, 
Vincent Kofman$^{4}$, 
Giulia Roccetti$^{5}$, 
Kenneth Goodis Gordon$^{6}$, 
Frans Snik$^{7}$, 
Theodora Karalidi$^{6}$, 
Victor Trees$^{8}$, 
Daphne Stam, 
Mary N. Parenteau$^{4}$}}

\affil{$^1$\small\it IRSOL, Universit\'a Svizzera italiana, Via Patocchi 57, 6605, Locarno, Switzerland}
\affil{$^2$\small\it University of Bern, Switzerland}
\affil{$^3$\small\it SETI Institute, USA}
\affil{$^4$\small\it NASA, USA}
\affil{$^5$\small\it ESO, Germany}
\affil{$^6$\small\it University of Central Florida, USA}
\affil{$^7$\small\it Leiden Observatory, Leiden University, the Netherlands}
\affil{$^8$\small\it Technical University Delft, Netherlands}




\begin{abstract}
{
Our Earth, being the only living planet that we know, provides us with clues that photosynthetic life-forms may be dominant on other exoplanets for billions of years. Spectropolarimetric signatures of the terrestrial photosynthetic life (PSLife) are well studied in the lab and remotely sensed with space and airborne instrumentation. An astonishing biosignature revealed by these measurements is an extremely strong linear polarization (tens \%) associated with broad absorption bands of biological pigments (biopigments) driving photosynthesis in various organisms. Also, unique circular-polarization signatures are associated with biopigments and other complex macromolecules as a sign of homochirality which is ubiquitous in terrestrial life forms. Thus, low-resolution spectro-  or multi-band polarimetry of exoplanets directly imaged at an unprecedented contrast using the HWO coronagraph is a novel opportunity for a robust discovery of life on exoplanets.
Here we propose to carry out two surveys and two follow-up observing programs. Survey 1 will identify potentially habitable planets (PHPs) through detection of atmospheres, clouds and liquid surface water (ocean) using linear polarimetry. Survey 2 will identify Living World (LW) candidates among PHPs by searching for strong linear polarization signatures associated with strong and broad absorption bands reminiscent of terrestrial biopigments. Follow-up program 3 will obtain multi-color surface maps of LWs, determine the distribution and abundance of alien photosynthetic organisms with exo-biopigments (exoBPs) and correlate their properties with the atmospheric and surface compositions. Follow-up program 4 will employ circular polarization to verify homochirality of exoBPs. This comprehensive approach aims at providing a quantitative answer to the ultimate question "Are we are alone in the Universe?". 
}
  \\
  \\
\end{abstract}

\vspace{2cm}

\section{Science Goal}
\label{sec:goal}

The fundamental question we aim to address in this HWO science case is: 
{\bf How common is photosynthetic life in the Universe?}
This broad question can be split up into several questions, such as:

    $\bullet$ How common are atmospheres and oceans on planets? 
    
    $\bullet$ How common are Living Worlds, in particular those with photosynthetic life? 
    
    $\bullet$ Do the abundance and type of photosynthetic life correlate with the properties of the planet and host star?
    
    $\bullet$ Is alien photosynthesis similar to that on Earth?   

In this paper we present a comprehensive path towards answering these questions using polarimetry combined with the Habitable Worlds Observatory (HWO) coronagraph capable of achieving an unprecedented planet-to-star flux ratio (contrast) of down to 10$^{-10}$, i.e., 0.1 parts per billion (ppb). The proposed observing programs are based on the current state-of-the-art theoretical models, laboratory and field measurements of terrestrial samples, and the planned capabilities of the HWO.

\subsection{Motivation}
\label{sec:motiv}

To address this fundamental goal, we propose to investigate potentially habitable planets (PHPs) and detect among them Living Worlds (LWs) that are populated by photosynthetic life (PSLife). Earth is so far the only LW known to us. With a huge diversity of life which emerged and evolved on Earth in the past and exists now, it provides us with a unique guidance for searching other LWs. In this section, we provide motivation for the goal and proposed approach.

\subsubsection{What are Living Worlds?}
\label{sec:lw}

Living Worlds are exoplanets exhibiting multiple undisputable bio- or techno-signatures.

Since the prerequisite of life (as we know it) is the presence of surface water, Earth-like potentially habitable planets (PHPs) are defined as rocky exoplanets in a circumstellar habitable zone (HZ) of around FGKM dwarfs \citep[e.g.,][]{Kopparapu2013}. Their size can be conservatively assumed in the range of 0.5--1.6 R$_{\rm Earth}$, or more broadly within 0.1--3 R$_{\rm Earth}$ (e.g., Habitable Worlds Catalog). The PHP atmospheric pressure and surface temperature allow for surface liquid water. The atmosphere may contain biogenic gases as a result of co-evolution of the geo- and bio- spheres \citep[e.g.,][]{CatlingClaire2005, VoroninBlack2007, Olson2018, Kasting2025}. 

LW is a PHP with an atmosphere and surface composition rich in elements and building blocks of life \citep[e.g.,][]{KitadaiMaruyama2018}, including the major elements C,H,O,N,S,P needed to sustain a habitable planet \citep[e.g.,][]{ Gaillard2021}. LWs are identified by a number of bio-/techno-signatures \citep[e.g.,][]{Kiang2018,Meadows2018,Schwieterman2018, Catling2018,Fujii2018,Walker2018,Berdyugina2018, BerdyuginaKuhn2019,Haqq-Misra2022}. Non-terran life “fingerprints" can be identified through molecular patterns and complexity \citep[e.g.,][]{Johnson2018}.

\subsubsection{Why focus on photosynthesis?}
\label{sec:ps}

On Earth, photosynthetic life (PSLife) emerged quite early, within the first billion of its history. Overall, photosynthetic organisms have been the most successful forms of life, which have currently achieved the largest biomass on Earth, and constituted the dominant life-form on Earth for billions of years \citep[e.g.,][]{DesMarais2000, DesMarais2002,Bar-On2018}. Therefore, we aim at detecting an extrasolar PSLife which is defined as organisms which are able to convert photonic energy into chemical energy.

Due to the abundant energy provided by the stellar radiation, photosynthesis may also evolve on PHPs throughout the Universe. Thus, detecting signatures of exoplanetary biological pigments, i.e., exo-biopigments (exoBPs), \citep[e.g.,][]{Kiang2007a,Kiang2007b,Berdyugina2016,Patty2022} together with other key biosignatures of LWs would be an indisputable discovery of extraterrestrial life.

\subsubsection{ What are the expected biosignatures?}
\label{sec:bs}

PSLife exploits many various biopigments to harvest and process photons. Thus, biopigments are key biosignatures of terrestrial PSLife. They efficiently capture and convert stellar light into chemical energy \citep{Scholes2011}. While chlorophyll-a is the primary pigment of cyanobacteria, algae and plants, there are up to 200 accessory and secondary (synthesized) biopigments, including other forms of chlorophyll, carotenoids, anthocyanins, phycobiliproteins, etc. These biopigments absorb almost all light in the visible range \citep[e.g.,][]{Kiang2007a}, which results in extremely broad absorption bands with a very high linear polarization \citep{Berdyugina2016}. Beyond Earth, PSLife may employ similar molecules for efficient light harvesting in spectral regions where useful photons are most abundant.

Homochirality (handedness) of terrestrial biopigments and other complex biomolecules is considered a universal and agnostic sign of life. For instance, amino acids and sugars in life forms are predominantly left- and right-handed, respectively. Thus, homochirality is a global biosignature, but its origin is still a mystery \citep[e.g.,][]{Frank1953, Blackmond2010, Laurent2021}. Detecting it in non-terrestrial organisms would provide an additional insight on life origin.

These two key photosynthetic biosignatures combined with other LW biosignatures (e.g., water and biogenic gases) detected on other planets (including those in the Solar system) will provide an indisputable detection of extraterrestrial life. 

\subsubsection{Why use polarimetry?}
\label{sec:pol}

Linearly and circularly polarized light reveals the presence of PSLife unambiguously and with a high contrast, and it helps characterize PHPs.

Terrestrial biopigments cause strong linear-polarization signatures in reflected light, up to several tens \%, associated with their broad absorption bands in the visible and near-infrared (NIR), where they are detected with a high contrast \citep{Berdyugina2016,Patty2022}. These uniquely strong linear-polarization signatures distinguish biopigment absorption bands from those of minerals and atmospheric gases \citep[e.g.,][]{Berdyugina2016}.

Homochirality of biopigments and other complex bio-molecules is detected in circularly polarized light \citep[e.g.,][]{Sparks2009b, Patty2019, Patty2021, Sparks2021}.

Polarization measurements can help in revealing macromolecular structures of exoBPs and inferring probable biochemistry and metabolism of organisms on exoplanets \citep[e.g.,][]{Patty2022}.

Polarimetry is also a powerful technique for discovering and characterizing PHPs. Linear polarization helps identify surface and atmospheric properties, such as ocean glint, Rayleigh scattering, cloud properties and surface composition, both abiotic and biotic \citep[e.g.,][]{Stam2008, Berdyugina2016, TreesStam2019, Sterzik2020, TreesStam2022, GoodisGordon2025a, Roccetti2025a}. 

Polarization signals constitute the only method to break degeneracies between albedo and size, as was demonstrated for asteroids \citep[e.g.,][]{Masiero2021} and cloud particles on Venus \citep[e.g.,][]{HansenHovenier1974}, and other Solar system planets. This is crucial for non-transiting exoplanets. 

Polarimetry helps increase the exoplanet contrast with respect to the stellar background and exozodi scattered light, the latter being probably the largest source of astrophysical noise in direct imaging \citep[e.g.,][]{Roberge2012}. However, when observed in linearly-polarized light, scattering from exozodi dust and exoplanets can be distinguished using polarimetric differential imaging \citep[e.g.,][]{Quanz2011}.

\subsubsection{What are the unique diagnostic capabilities of HWO?}
\label{sec:hwo}

A polarimetric multi-band instrument combined with a high-contrast coronagraphic imager can provide a major breakthrough in discovering life beyond Earth.

Linear polarimetry combined with high-contrast imaging and multi-band spectrophotometry will enable detection of  true PHPs with surface oceans (Section~\ref{sec:sc1}), robust discovery of PSLife and first maps of life colonies on LWs (Sections~\ref{sec:sc2}~and~\ref{sec:sc3}). Circular polarimetry will verify whether homochirality is omnipresent in living organisms beyond Earth (Sections~\ref{sec:sc4}).

The instrument required for implementing these science cases is similar to that on the Roman CGI which will be first to demonstrate high-contrast coronagraphic imaging polarimetry at the contrast of at least 0.1 parts per million (ppm) which has not been yet achieved with large ground-based telescopes \citep{Groff2021}. The Roman CGI will help fine-tune the HWO observing programs proposed here. 

We note that the high-resolution spectro-polarimeter POLLUX proposed earlier for LUVOIR and now for HWO is intended to be used without the coronagraph.

\subsection{Topics Related to the Astro2020}
\label{sec:astro}

The scientific goal addressed in this Science Case is within the framework of the Astro2020 theme "Worlds and Suns in Context", in particular the Priority Area "Pathways to Habitable Worlds", which recommends {\em efforts to identify habitable Earth-like worlds in other planetary systems and search for the biochemical signatures of life, that will play a critical role in determining whether life exists elsewhere in the universe. (nap.nationalacademies.org)}

The science case proposed here is within the Discovery Area "The Search for Life on Exoplanets". It is 
directly related to the science question Q4 and also contributes to Q2 and Q3, which were formulated by the Panel on Exoplanets, Astrobiology, and the Solar System (EAS):

$\bullet$ Q2: What are the properties of individual planets, and which processes lead to planetary
diversity?

$\bullet$ Q3: How do habitable environments arise and evolve within the context of their planetary
systems?

$\bullet$ Q4: How can signs of habitable life be identified and interpreted in the context of their
planetary environments?





\section{Science Objective}
\label{sec:obj}

To achieve understanding how common photosynthetic life on exoplanets, the following overarching objective is addressed here:
{\bf Search for, characterize and determine the occurrence rate of habitable exoplanets and diverse photosynthetic life using multi-wavelength coronagraphic imaging polarimetry with HWO.}

This objective is addressed in four science cases (SC) which provide the basis for corresponding science observing programs with HWO (Fig.~{fig:scheme}).
Specific objectives, observing modes and expected outcomes of these SCs are summarized below.

\begin{figure*}[ht!]
    \centering
    \includegraphics[width=1\textwidth]{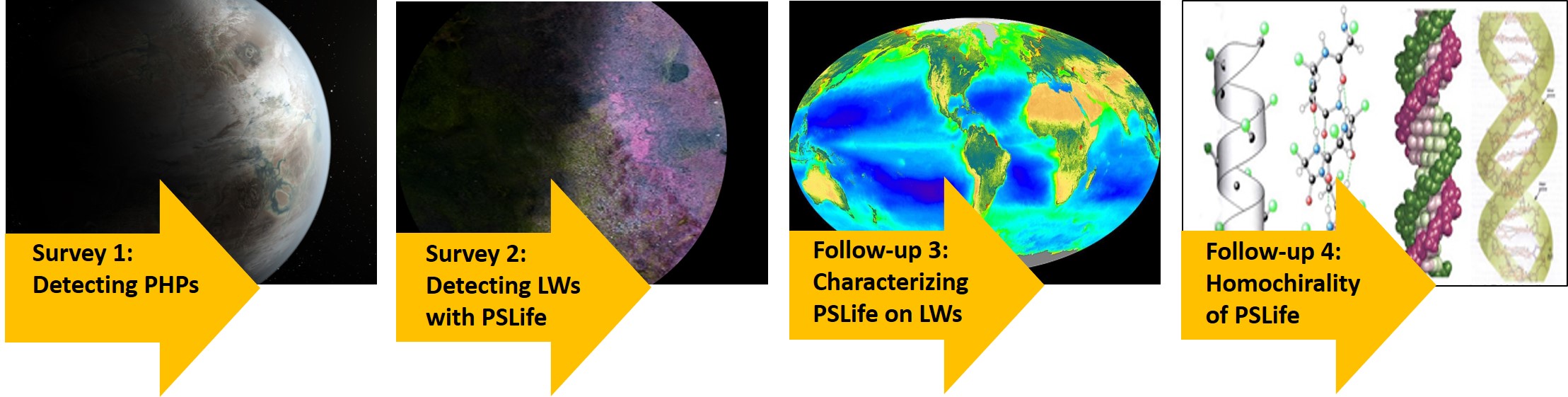}
    \caption{A schematic of four observing programs (science cases) proposed in this HWO science case to discover and characterize life beyond Earth.}
    \label{fig:scheme}
\end{figure*}

{\bf SC1 / Survey 1: Detect PHPs.} \textit{Linear-polarimetry} multi-wavelength survey of HZ rocky planets will be carried out to assess the likelihood of their habitability through detection of an atmosphere, clouds, and surface water.

    Observing mode: 2--3 visits per target at 2--3 selected orbital phases and any planetary axial rotation phases.
 
    Outcomes: detection of moderately or strongly polarized signatures with specific wavelength dependencies, that are indicative of atmospheres, clouds, and surface water, i.e., PHPs.

{\bf SC2 / Survey 2: Detect LWs.} \textit{Linear-polarimetry} multi-wavelength survey of the PHPs with a high habitability likelihood identified in Survey 1 will be carried out to reveal the possible presence and types of exoBPs, i.e., LWs with PSLife.

    Observing mode: single visit per target, near an orbital quadrature (half-disk illumination), with measurements at 5--10 evenly distributed planetary axial rotation phases.
 
    Outcomes: first detections of LW candidates with PSLife, estimates of the occurrence rate of such LWs among PHPs and HZ rocky planets, identification for each LW candidate of the axial rotation phases with a significant presence of exoBPs.

{\bf SC3 / Follow-up 3: Characterize LWs.} \textit{Linear-polarimetry} multi-wavelength follow-up observations of the LW candidates identified in Survey 2 will be carried out to obtain LWs' multi-color spatially-resolved surface maps, evaluate surface area fractions covered by land, ocean and photosynthetic organisms, and determine their chemical composition.

    Observing mode: 10--20 visits per target at selected orbital phases, each covering one planet rotation at 5--10 evenly distributed phases.
    
    Outcomes: first surface maps of alien LWs, robust detection of PSLife on exoplanets, relations of exoBP abundance and types with abiotic surface constituents and atmospheric compositions.

{\bf SC4 / Follow-up 4: Verify homochirality on LWs.} \textit{Circular-polarimetry} multi-wavelength follow-up observations of one or more LWs with the highest abundance of exoBPs identified in Follow-up 3 will be carried out to search for a homochirality signature as a universal and unambiguous signature of life.

    Observing mode: 2--3 visits per target at selected orbital and axial rotation phases to accumulate a significant circularly-polarized signal associated with exoBPs.
    
    Outcomes: unambiguous confirmation of life detection(s) on exoplanets, identification of potential biospheres containing a large abundance of asymmetric macromolecules (biopolymers).

\subsection{SC1 / Survey 1: Detect PHPs}
\label{sec:sc1}


This science case (Survey 1) aims at observing HZ rocky planets and identifying PHPs by detecting moderately or strongly linearly polarized signatures with specific wavelength dependencies that are indicative of atmospheres, clouds, and surface water, i.e., prerequisites of Earth-like life.

The limited photon flux from directly imaged planets in the visible to near-infrared necessitates the optimal use of potential informational content of the photons. Linear polarization may provide several critical additions to understand the potential habitability of a rocky planet by breaking critical degeneracies, such as unambiguous estimates of planetary size and albedo, unambiguous detections of the presence of liquid water in the form of oceans and clouds, and independent confirmations of detections of biosignature gases. 

\begin{figure*}[ht!]
    \centering
    \includegraphics[width=1\textwidth]{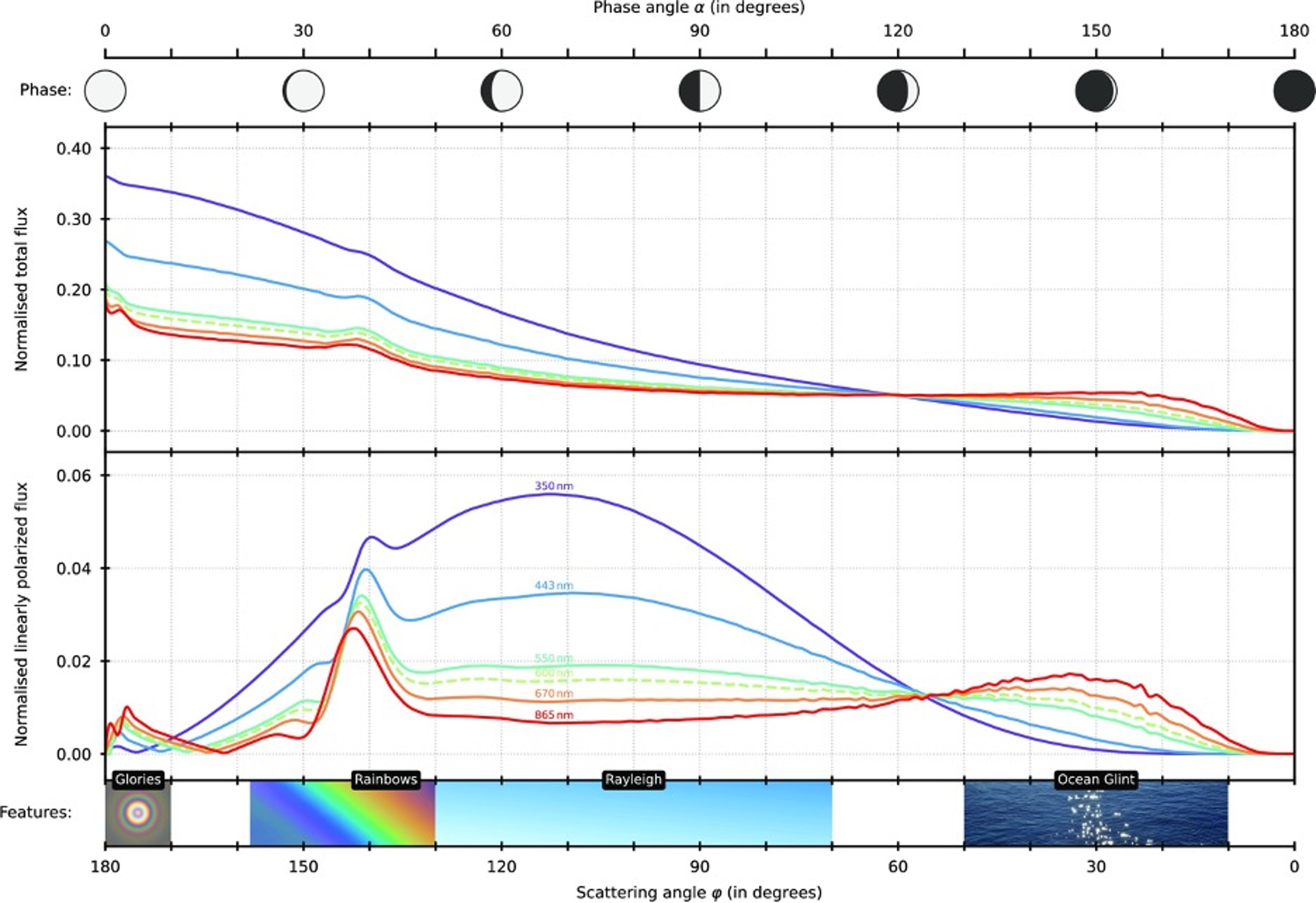}
    \caption{
    Phase curves of the total flux (top) and the normalized linearly polarized flux (bottom) of a cloudy planet that is completely covered with a wavy ocean and that has an Earth-like atmosphere. The degree of linear polarization is the ratio of the polarized flux to the total flux for each wavelength and scattering angle (not shown).  The different colors indicate different wavelengths (see the labels). The wind speed (which determines the wave height, following \citet{CoxMunk1954} is 7 m/s, and the clouds consist of spherical water droplets. The cloud fraction is 50\%. While the cloud pattern is patchy, in these simulations, the glint is cloud-free. The total fluxes are normalized such that they equal the planet’s geometric albedo at zero phase angle (when the planet disk is fully illuminated, phase=0). The panel at the bottom indicates the ranges of scattering angles (180 - planetary phase angle) where angular features due to the glory, the rainbows, Rayleigh scattering, and ocean glint leave traces in the curves. Adapted from \citet{Vaughan2023}.
    }
    \label{fig:sc1}
\end{figure*}

The degree of linear polarization of starlight reflected by an exoplanet is very sensitive to the optical properties of the atmosphere, cloud particles and surfaces, and it also depends on the wavelength and scattering angle \citep[e.g.,][]{HansenTravis1974}. Consequently, the phase curves in polarized light have typically more distinct peaks from different scattering sources than in the total (polarized + unpolarized) light \citep[e.g.,][]{Berdyugina2008, Berdyugina2011, Stam2008, FluriBerdyugina2010, Berdyugina2016b, TreesStam2019, TreesStam2022, Vaughan2023, GoodisGordon2025a, Roccetti2025a, Roccetti2025b}. As an example, Fig.~\ref{fig:sc1} presents planetary phase curves of the total and polarized flux of disk-integrated light reflected by an Earth-like exoplanet. 

{\bf Atmosphere.} 
Rayleigh scattering by gas in the atmosphere is strongly polarized at a phase angle of 90$^\circ$. This is an important indicator of the presence of an atmosphere, constraining atmospheric pressure, and it is the only way in which the HWO will be able to detect spectroscopically inactive gasses such as N$_2$, H$_2$ or He, as well as clouds and hazes. Also the O$_2$ A-band polarization peaks above the surrounding continuum polarization due to Rayleigh scattering \citep[e.g.,][]{Stam2008, Sterzik2019, Roccetti2025b}.

{\bf Ocean glint.} 
Light reflected by the ocean results in an ocean glint, whose size increases with decreasing wind speed (i.e., the inclination of the ocean waves) and increasing planetary phase angle \citep[e.g.,][]{CoxMunk1954, WilliamsGaidos2008, Zugger2010, Kopparla2018}. The ocean glint strongly polarizes the light due to Fresnel reflection, which is best visible in the red to near-infrared wavelengths (Fig.~\ref{fig:sc1}), where the opacity of the atmosphere due to Rayleigh scattering is relatively small compared to that at shorter wavelengths \citep[e.g.,][]{Zugger2011, TreesStam2019, Vaughan2023, Roccetti2025b}. 

A variable cloud cover on the planet (and weather patterns) may complicate the interpretation of measured phase curves. However, using the spectropolarimetry of the disk-integrated planet, the ocean glint can be detected at a single planetary phase angle, if the glint is not fully covered by local clouds \citep{TreesStam2022, Roccetti2025a}. In particular, the degree of polarization is increased in the continuum (i.e., wavelengths at which atmospheric gases do not absorb), and decreases in gaseous absorption bands (at which the ocean surface is invisible).

The presence of land may complicate the detection of ocean glint  \citep[e.g.,][]{Groot2020, Roccetti2025a}. In unpolarized flux, models with pure ocean and land-ocean mixture are nearly indistinguishable, but they exhibit significant differences in polarization at phase angles greater than 90$^\circ$. Earthshine spectropolarimetric observations at 700 nm validate these models \citep{Sterzik2019, Sterzik2020}.

\begin{figure*}[ht!]
    \centering
    \includegraphics[width=0.75\textwidth]{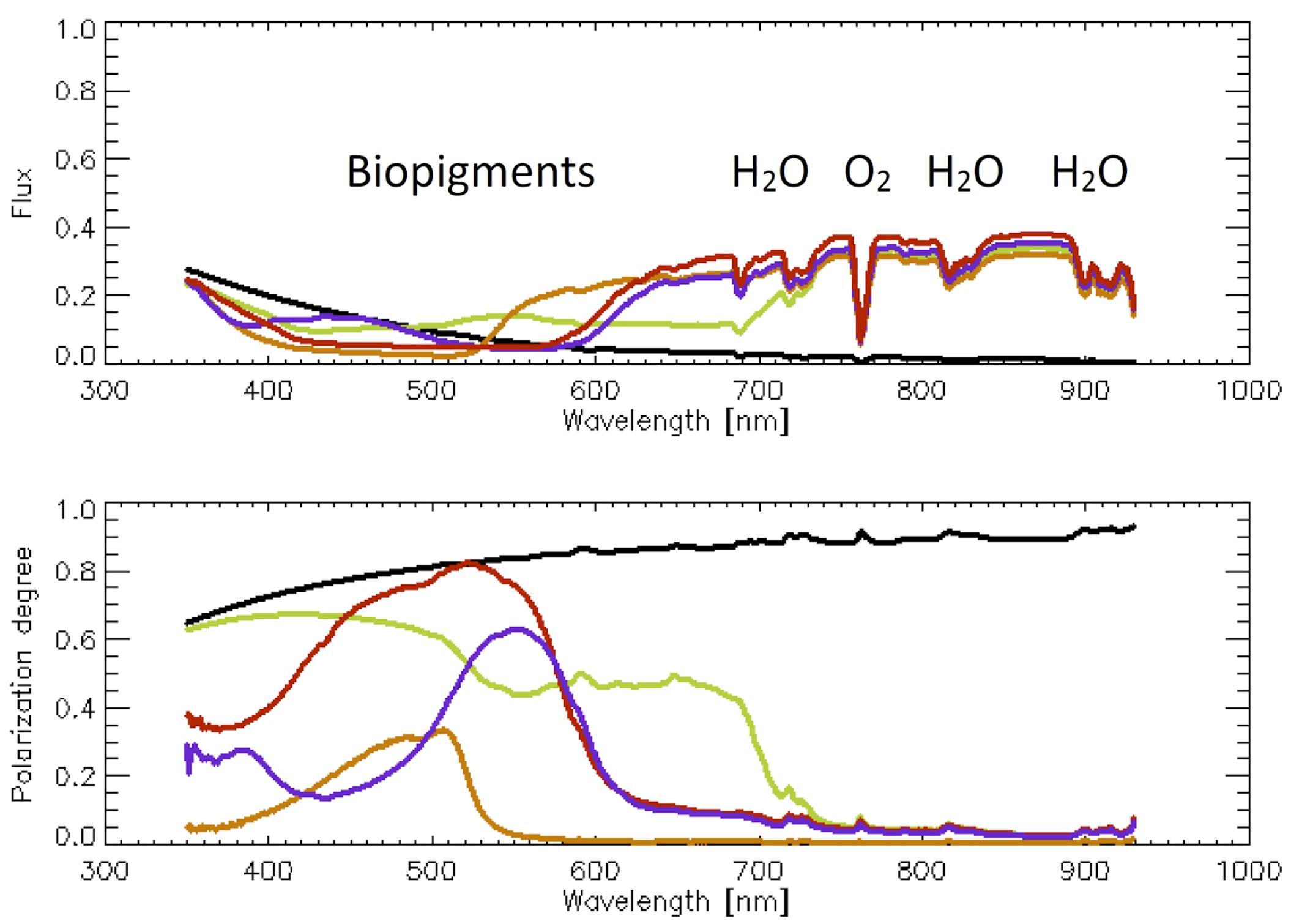}
    \caption{
    Modelled reflectance spectra (top) and linear polarization degree spectra (bottom) for exoplanets with a cloud-free Earth-like atmosphere and 80\% surface coverage by phototrophic pigmented organisms and 20\% ocean surface coverage (visible hemisphere only). Green, red, yellow, and purple curves correspond to planet models with organisms predominantly reflecting such colors. The black curve represents a planet with a water ocean only (flat Fresnel reflection, no ocean glint), the narrow absorption bands at the 700--950 nm polarized at a few percent are atmospheric water and oxygen bands \citep{Stam2008}. From \citet{Berdyugina2016}.
    }
    \label{fig:sc2}
\end{figure*}

{\bf Clouds and Rainbows.}
Polarization offers more powerful diagnostic capabilities than reflectance (albedo) alone for distinguishing between cloud-free and cloudy exoplanets \citep[e.g.,][]{Stam2008, RossiStam2017, Roccetti2025b}. Both the spectral slope and polarization degree are highly sensitive to the presence of clouds, while features like the cloudbow provide insights into cloud microphysical properties \citep{Sterzik2020, Roccetti2025a}. Unlike reflectance, which becomes less informative at larger phase angles, polarization becomes increasingly sensitive, especially to cloud characteristics such as particle composition and structure \citep[e.g.,][]{HansenHovenier1974}, including the presence of ice crystals \citep[e.g.,][]{Karalidi2012, Emde2017, Roccetti2025a}, and also for probing cloud depth \citep[e.g.,][]{Fauchez2019}.

Cloud backscattering optical effects, such as rainbows and glories, produce characteristic linear polarization signatures. With sufficient phase curve sampling, the peak angle of the rainbow feature can be analyzed to provide information about the primary species constituents of the condensates \citep[e.g.,][]{Bailey2007,  Karalidi2011, Roccetti2025a}. The relative strength of the rainbow compared to the Rayleigh scattering can provide a constraint to the atmospheric depth of the cloud top altitude \citep[e.g.,][]{Gordon2023}. 

\subsection{SC2/Survey 2: Detect Living Worlds}
\label{sec:sc2}


This science case (Survey 2) aims at first detections of LW candidates with PSLife, determining the occurrence rate of such LWs among PHPs and HZ rocky planets, and identifying candidate targets for detailed follow-up studies. This will be achieved by detecting exoBPs in linearly polarized light at multiple wavelengths reflected from the PHPs identified in the SC1 (Survey 1).

Early polarimetric measurements of light reflected from green leaves \citep[e.g.,][]{Pospergelis1969, Wolstencroft1974} revealed a high linear polarization caused by the chlorophyll absorption. Subsequently, the chlorophyll’s absorption bands, red edge and green bump occurring in terrestrial vegetation were suggested as a potential biosignatures on exoplanets \citep[e.g.,][]{WolstencroftBreon2005, Kiang2007a, Stam2008, Berdyugina2016}. They were also searched for in the Earthshine polarized spectrum reflected from a dark Moon surface as a proxy to exoplanets \citep[e.g.,][Roccetti et al. 2025c, submitted]{Sterzik2012, Sterzik2019, Emde2017}.

In addition to chlorophyll, other terrestrial biopigments were also found to cause large-degree linear-polarization signatures, up to several tens \%, associated with their absorption bands when observed at certain scattering angles between incident and outgoing beams \citep{Berdyugina2016, Patty2022}. The maximum absorption, polarization degree and its phase-dependence on scattering angles are characteristics of particular biopigments. In particular, their polarization signals are stronger and angular dependence is different from those produced by minerals due to the Umov effect. These characteristics of biopigments, which cannot be obtained through spectroscopy only, together with the central wavelength and the width of the polarized signatures, serve as markers for particular molecular constituents of the biopigments. 

The model example shown in Fig.~\ref{fig:sc2} \citep{Berdyugina2016} demonstrates the overall high sensitivity of polarimetric measurements to various assemblies of terrestrial biopigments, in contrast to the flux measurements at the same wavelengths. In the visible part of the spectrum, the flux is reduced by about 20--30\% due to biopigment absorption, while their polarization reaches 30–80\%. The NIR part of the spectrum ($>$700 nm) indicates the presence of water and oxygen in the terrestrial atmosphere. The expected maximum polarization degree for the organisms containing chlorophyll (green curve in Fig.~\ref{fig:sc2}) seen through the Earth-like atmosphere agrees well with the measurements by the French (CNES) POLDER satellite of the cloud-free forest canopy: 60\% at 443 nm, 28\% at 670 nm and 7\% at 865 nm \citep{WolstencroftBreon2005, Wolstencroft2007}. This comparison also illustrates that polarization cancellation due to different angles of leaf surfaces with respect to the light source and viewing angles is not a major problem, but this effect should be approximated for more precise modelling. Because of Rayleigh scattering in the Earth atmosphere the reflected light gains a contribution in the blue part. This is exactly what is seen from space. In the UV, the polarization may reduce again due to multiple Rayleigh scattering.


This example illustrates at which wavelengths various biopigments can be detected. The absorption depth and polarization peak depend on the surface percentage covered by biopigments, ocean, bare land and clouds. The polarization degree also varies with the angles with respect to the stellar illumination direction and the observer. In particular, clouds dilute the signatures of both biopigments and atmosphere \citep{Berdyugina2016}. A completely cloudy atmosphere will obviously disguise the presence of biopigments (and everything else) on the planetary surface, while a small cloud coverage of around 20\% will only marginally reduce polarization effect. Thus, clouds are the most disturbing factor in detecting surface biosignatures. However, weather variability and short life-time of many cloud types should assist in successful detection if a planet is monitored long enough to reveal long-lived features on the surface \citep{BerdyuginaKuhn2019}.

\subsection{SC3/Follow-up 3: Characterize Living Worlds}
\label{sec:sc3}


This science case (Follow-up 3) aims at robust detection of photosynthetic life on exoplanets by obtaining first surface maps of alien LWs, determining surface area fractions covered by land, ocean and photosynthetic organisms (abundance of PSLife), and identifying their chemical composition.

Seeing oceans, continents, quasi-static weather and seasonal patterns, life colonies and even artificial structures on exoplanets is necessary for a robust detection and characterization of life outside the Solar system.

Time series of measurements of the light reflected from an exoplanet (light-curves) contain information on both longitudinal and latitudinal structures and can provide exoplanet surface albedo maps at different wavelengths. Model solutions obtaining albedo maps, along with the spin-orbital parameters, from reflected flux or polarization light-curves have been previously demonstrated for angularly unresolved planets, moons and asteroids in the Solar System, and for various types of exoplanets, from hot Jupiters to Earth-like \citep[e.g.,][]{Russell1906, Guthnick1906, Morrison1975, Buie1997, Kaasalainen1992, Carbognani2012, Cowan2009, FluriBerdyugina2010, KawaharaFujii2011, FujiiKawahara2012, Schwartz2016, Cowan2017, Lustig-Yaeger2018, Farr2018, BerdyuginaKuhn2019, Aizawa2020, Kawahara2020, KawaharaMasuda2020, AsensioRamosPalle2021, Kuwata2022, Teinturier2022}. These and other works differ by assumptions and by numerical approaches to forward modelling and deconvolution (inversion) of light-curves, which affect properties of inferred ‘best’ solutions and their uncertainties. Resulting models vary from simple 1D back-projection “maps” to 2D principal-component maps, and more realistic 2D pixel-wise inferred multi-colour maps. Assumptions differ also about star-planet geometries and compositions of the planetary surface and atmosphere; see complementary reviews by \citet{Berdyugina2019} and \citet{CowanFujii2020}.

\begin{figure*}[ht!]
    \centering
    \includegraphics[width=0.35\textwidth]{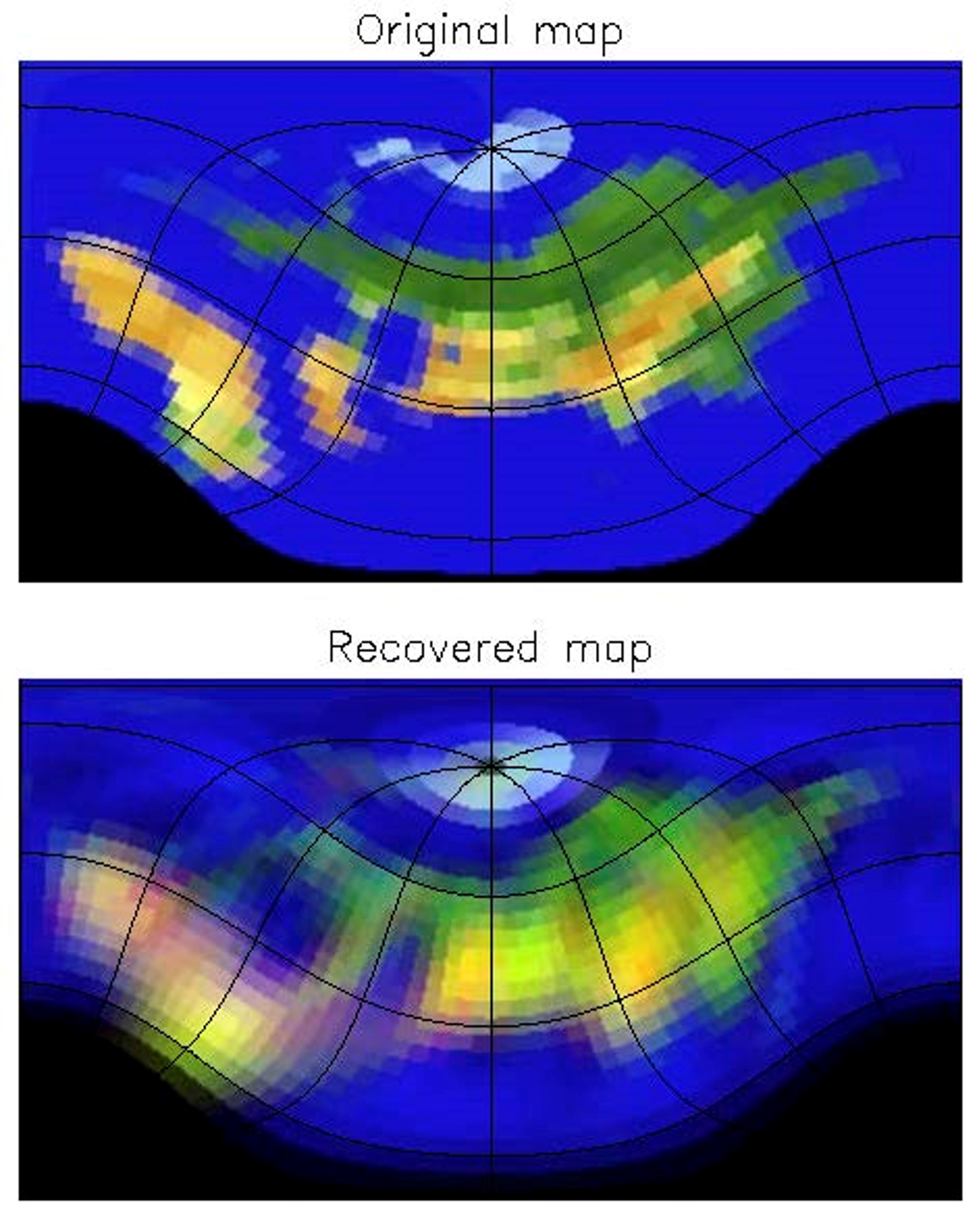}
    \includegraphics[width=0.6\textwidth]{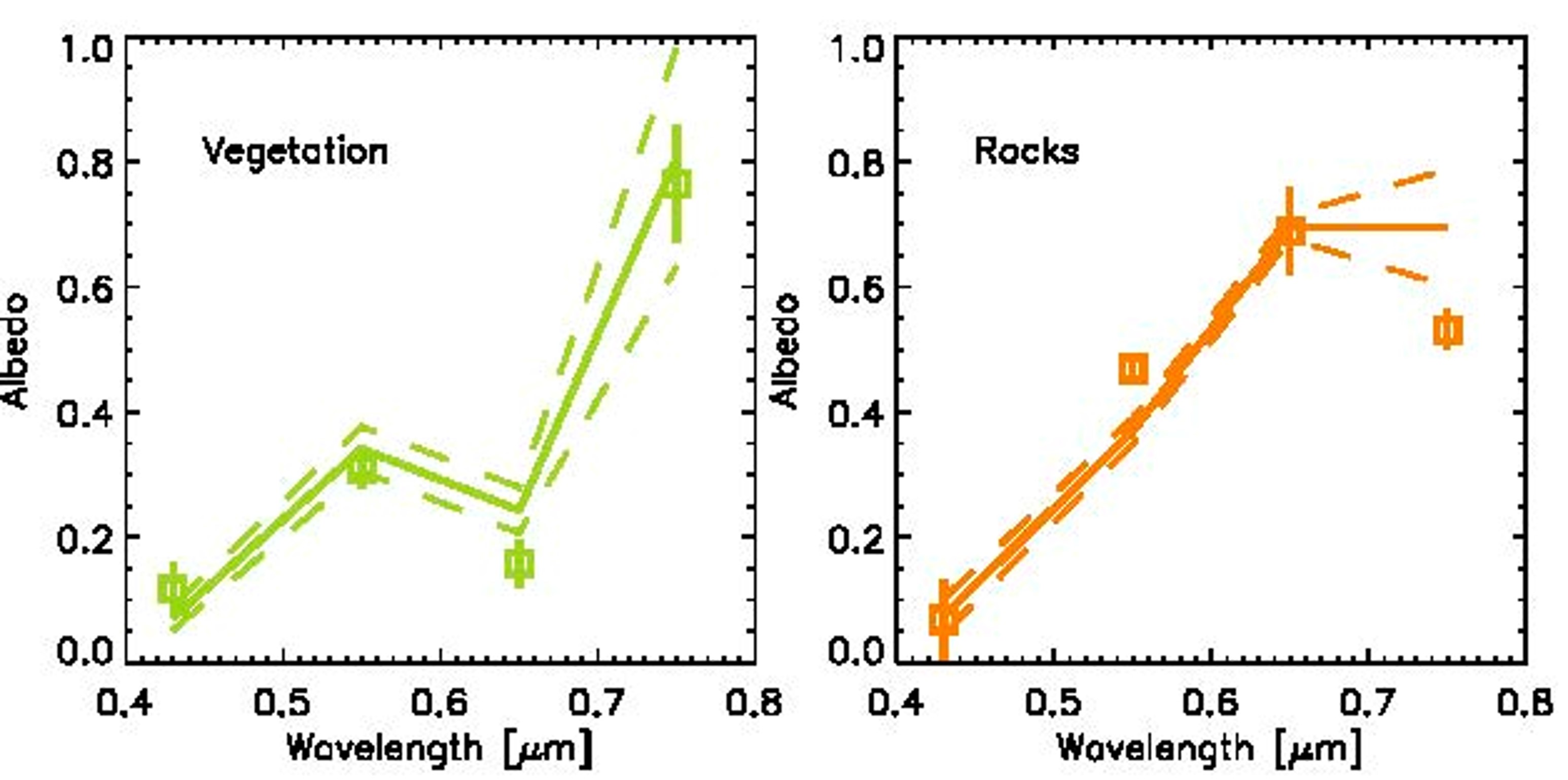}
    \caption{
    {\em Left}: A composite albedo map of an Earth-like exoplanet. The original map is based on a composite Earth image using space data with parts of the terrestrial continents arranged in a likely pattern. The recovered map is inferred from synthetic observations in four ONIR bands using the EPSI inversion technique. The achieved subcontinental spatial resolution allows for a robust identification of deserts, green vegetation, a polar ice-cap, and deep ocean. {\em Right}: Four-wavelength spectra extracted from the resolved desert and vegetation areas. The retrieved spectra (symbols with error bars) fit well the input model spectra (solid lines with the standard deviations within the model surface area shown with dashed lines). Adapted from \citep{BerdyuginaKuhn2019}.
    }
    \label{fig:sc3}
\end{figure*}

When light curves are measured in several wavelength passbands (optical to NIR, hereafter ONIR) and during both the rotational and orbital exoplanet periods, they can provide quite realistic “photographic” views of distant exoplanets, analogous to multi-spectral images of Earth from space (Fig.~\ref{fig:sc3}). Successful inversions of synthetic data for Earth, other Solar system planets and moons, simulated exoplanets with Earth-like life and artificial megastructures have been demonstrated using the Exoplanet Surface Imaging (EPSI) technique by \citet{BerdyuginaKuhn2019}, as well as by \citet{Kawahara2020} for Earth using a different numerical technique.

It was shown that even sparse, low signal-to-noise ratio (SNR) observations can clearly reveal the presence of the ocean and land areas in surface maps, mostly due to a large albedo difference of the ocean and land areas in the ONIR wavelengths, which is typical on Earth. Thus, the surface fractions of the land and ocean (the ocean-to-land ratio) can be directly deduced from the maps with a high degree of confidence. 

Furthermore, the surface composition can be determined from multi-wavelength spectral and polarimetric signatures which are spatially resolved in 2D maps (per surface resolution element). For example, for higher SNR and cadence, the EPSI technique can detect land areas down to 2\% of the total exoplanetary surface area (Fig.~\ref{fig:sc3}). On Earth, a large fraction of areas of this size is frequently covered by rather homogeneous substances, e.g., sands (Sahara, Australia), rocks (large mountain ranges in Asia and the Americas), and vegetation (Siberia, Amazon). In addition, large snow fields (Arctics) and polar ice-caps (Antarctica) can be clearly recognized. Such retrieval methods can be tested with data from the LOUPE/PEARL instrument \citep{Klindzic2021} that is currently being developed for time-resolved spectropolarimetric measurements of the Earth-as-an-exoplanet from the lunar surface/orbit.

Identifying spectral and polarization signatures of exoBPs within surface areas of 2–10\% significantly increases our chances to detect life-forms similar to those which are dominant on Earth, i.e., photosynthetic organisms. This is because of an increased surface fraction (concentration) of organisms covering such areas, as compared to that in the integrated light from an exoplanet (Survey 2, Section~\ref{sec:sc2}), which will provide candidates for the Follow-up 3 observations. In addition, the exoplanetary orbital and rotation axis parameters (obliquity and azimuth), atmospheric composition and the ocean-to-land surface ratio, determined simultaneously with the surface multi-spectral maps can inform us about possible seasons and climates on exoplanets.

\begin{figure*}[ht!]
    \centering
    \includegraphics[width=0.75\textwidth]{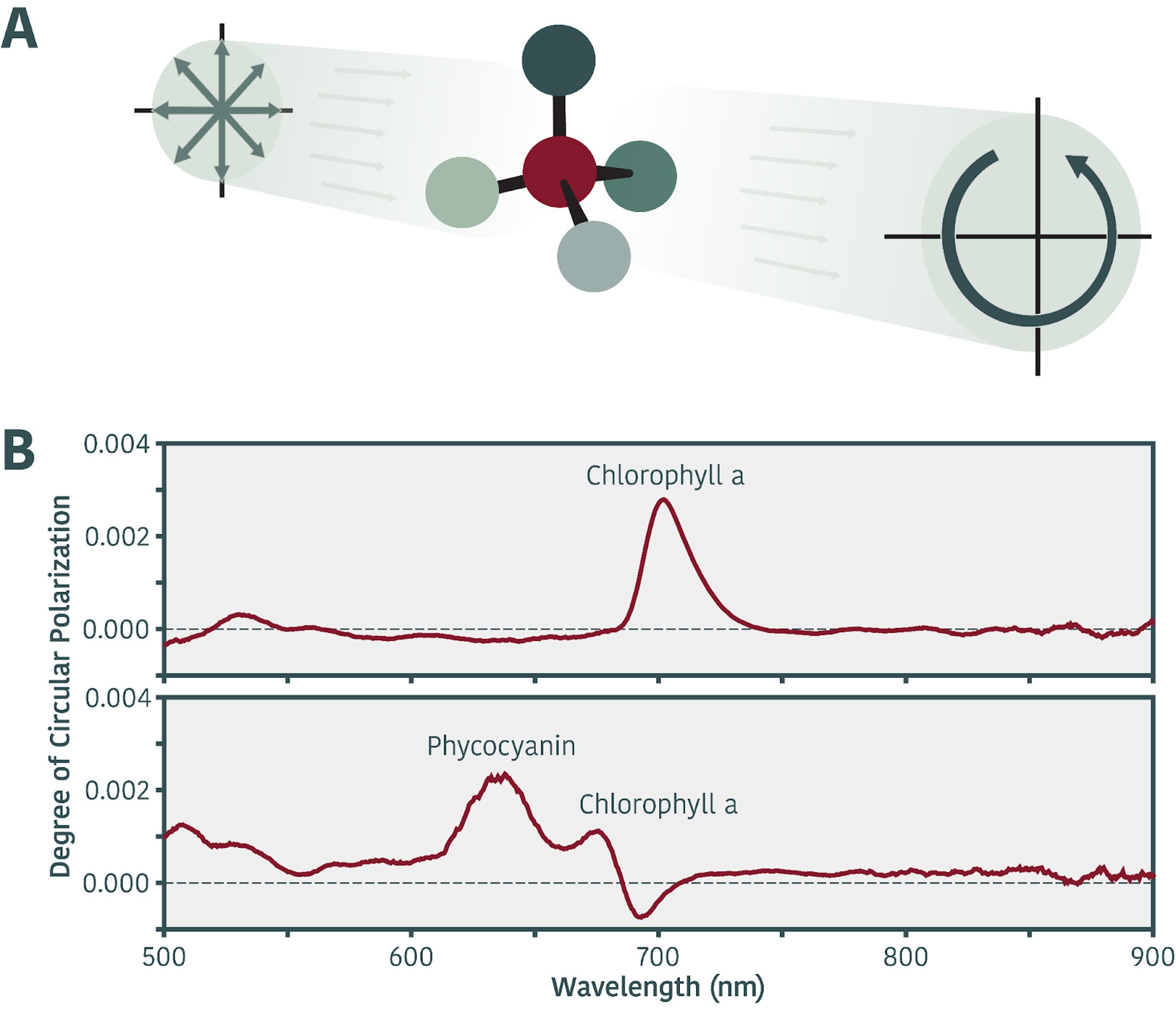}
    \caption{
    {\bf A}: Illustration of unpolarized light interacting with a homochiral macromolecule which induces partial circular polarization. {\bf B}: Examples of circular polarization spectra for a leaf in reflection (upper panel) and a cyanobacteria culture in transmission (lower panel). The data are unpublished measurements by J.~Grone (priv. comm.).
    }
    \label{fig:sc4}
\end{figure*}

Cloud and surface variability caused by weather and seasonal changes remain common issues in all map inversion techniques applied to examples based on Earth data. One solution is to employ inversions of multiwavelength light-curves. For instance, near-infrared (800--900\,nm) and blue (400--500\,nm) data improve reconstruction of the water/land distribution, except for at high latitude regions persistently covered by clouds and snow \citep{KawaharaFujii2011}. Also, surface albedo maps with a time-averaged cloud cover and time-dependent partial maps with seasonal variations in surface vegetation and ice cover can be obtained \citep{BerdyuginaKuhn2019}. Model reconstructions of large-scale evolving albedo patterns due to clouds demonstrated that heavily clouded LWs may remain a mystery \citep{Kawahara2020,AsensioRamosPalle2021}. Nevertheless, characterizing large-scale cloud patterns and properties of their particles can inform us about global processes on PHPs and LWs. Additional model studies are needed to further improve disentangling different components of the exoplanetary surface and atmosphere.

\subsection{SC4/Follow-up 4: Verify honochirality on LWs}
\label{sec:sc4}


This science case (Follow-up 4) aims at unambiguous confirmation of life detection(s) on exoplanets by identifying potential biospheres containing an abundance of asymmetric biopolymers. This is achieved by measuring circular polarization in light reflected by surfaces of LWs with the highest abundance of exoBPs identified in the Follow-up 3 program (Section~\ref{sec:sc3}). Detecting circular polarization as a homochirality signature would provide a strong indicator of a biosphere-dominated surface.

Terrestrial biochemistry is fundamentally dependent on chiral molecules, whose functionality underpins the complex macromolecules essential for life. Unlike abiotic chemistry, which produces equal amounts of both mirror-image forms (enantiomers), biological systems almost exclusively employ a single enantiomer. This strict preference for one chiral form of a specific macromolecule drives the phenomenon of homochirality, a near-absolute dominance in abundance of one chiral form which is a hallmark of biosphere-dominated environments. Homochirality ensures the orderly assembly and functionality of biopolymers. In its absence, the construction of such molecules would be chaotic, rendering them dysfunctional and incompatible with life’s metabolic processes \citep[e.g.,][]{Blackmond2010}.

On Earth, a very common example of abundant asymmetric biopolymers is displayed by chlorophyll molecules organized in chiral pigment-protein complexes. 
The asymmetry of the macromolecules used in photosynthesis endows them with a specific response to light, inducing fractional circular polarization upon reflection of the incident unpolarized light from the host star (Fig.~\ref{fig:sc4}A). These circular polarization features can be observed as characteristic, narrow-banded spectral peaks, correlating with the pigment absorption bands \citep[Fig.~\ref{fig:sc4}B,][]{Patty2018}. In addition, the circular polarization resulting from chiral molecules is probably relatively insensitive to alterations in phase angle \citep{Patty2022}.

\begin{figure*}[ht!]
    \centering
    \includegraphics[width=0.75\textwidth]{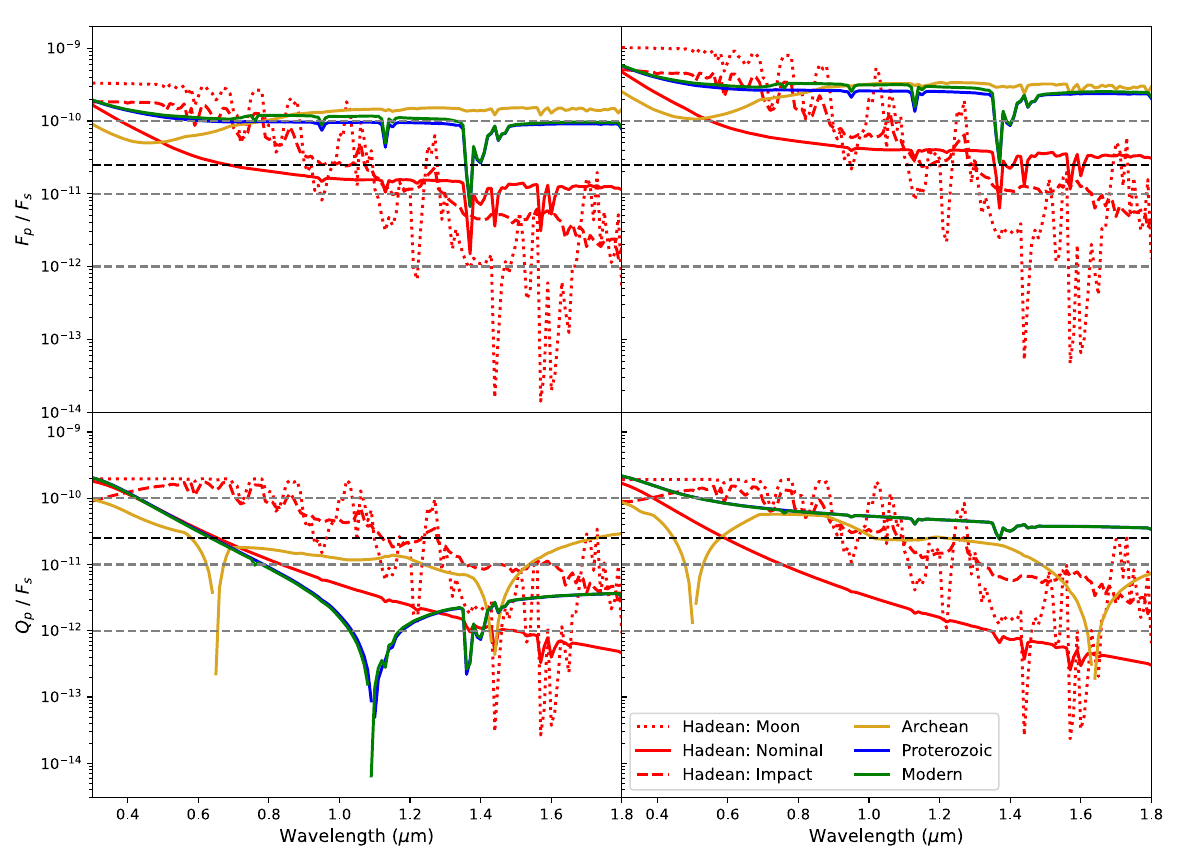}
    \caption{
    Unpolarized (top row) and polarized (bottom row) contrast ratios as function of wavelength for the Earth at six points throughout its geologic evolution. These contrast ratios assume the planets are in circular edge-on orbits at a distance 1 AU from the host star (assumed to be an evolving Sun). Note the distinguishing spectral features of different species across the different epochs, as well as a detectable VRE in the Modern Earth model (green lines). Adapted from \citet{GoodisGordon2025a}.
    }
    \label{fig:par1}
\end{figure*}

Circular polarimetric measurements of various minerals, including those on the surface of Mars, indicate the general absence of narrow-banded nonzero signals. This attests to the general lack of false-positive scenarios and abiotic mimicry \citep[e.g.,][]{Pospergelis1969, Sparks2009a, Sparks2009b, Sparks2012, Patty2019}.

Considering a relatively low circular polarization signature due to homochirality of macromolecules participating in photosynthesis, $\leq$1\% \citep[e.g.,][]{Patty2019, Sparks2021}, the expected signal from a cloud-free rocky exoplanet with a partial surface coverage by biomolecules would be $\leq$0.1\%. This may not be possible to detect with the HWO for rocky planets in HZ of G- and K-dwarfs, but it is still worth considering this follow-up observing program for late M-dwarfs and/or for a second-generation of the HWO instrumentation. 

\section{Physical Parameters}
\label{sec:par}

Table~\ref{tab:par} presents possible scenarios for detections and non-detections of PHPs and LWs with PSLife. The assumed numbers are proposed as hypotheses to be tested for given parameters. The challenges and paths for advancing the detections, identifications and interpretations are briefly commented in this Section for each physical parameter within the four observing programs. 

\subsection{Habitability likelihood of HZ rocky exoplanets (Survey 1)}
\label{sec:par1}

To assess the likelihood of habitability of HZ rocky exoplanets, we aim at detecting atmospheres, clouds, and surface water, which are characteristics of Earth-like PHPs. The PHP occurrence rate will be determined as a ratio of the number of PHP detections to the number of qualified visit observations of HZ rocky exoplanets.

\subsubsection{Atmospheres}
\label{sec:par1.1}

The presence and composition of atmospheres can be revealed through detection of absorption and scattering polarization by gas-state molecules. 

So far, only a few HZ rocky exoplanets could be observed and for which atmospheres were detected using the transit method. Considering that Earth's atmosphere has evolved through time and that other Earth-like planets may have yet other constituents, detecting both absorption and polarization is important for confident identification of specific molecules. For example, Figure~\ref{fig:par1} shows models with the wavelength dependence of contrast ratios in total flux and linear polarization for various Earth-through-time scenarios \citep[][]{GoodisGordon2025a}.

Example scenarios to be tested for this parameter:

$\bullet$ Incremental Progress (Enhancing) would be detecting atmospheres in at least half of the HZ rocky exoplanets. 

$\bullet$ Substantial Progress (Enabling) would be detecting atmospheres in about two thirds of the observed HZ rocky exoplanets.

$\bullet$ Major Progress (Breakthrough) would be detecting atmospheres in the majority of the observed HZ rocky exoplanets, e.g., 90\% of those.

\subsubsection{Liquid surfaces} 
\label{sec:par1.2}

Contrary to methods that search for the ocean glint in the total flux, \citep[e.g.,][]{Robinson2010}, the dips in the degree of linear polarization spectra can be searched for at intermediate phase angles where the planet-star separation is the largest, e.g., $\sim$90$^\circ$ for circular orbits \citep[e.g.,][]{TreesStam2022, Roccetti2025b}. The specific ocean-glint characteristics are the wavelength-dependent phase-curve crossing and dips (instead of peaks) in the degree of linear polarization spectrum (Fig.~\ref{fig:sc1}). This helps to distinguish the ocean glint from cloud reflection at large phase angles in the total (polarized + unpolarized) flux, because the latter hardly polarize the light due to multiple scattering by cloud droplets \citep[e.g.,][]{Stam2008, TreesStam2019, TreesStam2022, Roccetti2025a}. In addition, the polar caps of a planet can mimic the glint in the total flux due to the ‘latitude-albedo effect’ \citep[e.g.,][]{Cowan2012}. Measurements in polarized light break the degeneracy between ocean glint and these false-negative ocean glint features, allowing for a unique identification of an exo-ocean.

Example scenarios to be tested for this parameter:

$\bullet$ Incremental Progress (Enhancing) would be detecting oceans on quater of the observed HZ rocky exoplanets. 

$\bullet$ Substantial Progress (Enabling) would be detecting oceans on third of the observed HZ rocky exoplanets. 

$\bullet$ Major Progress (Breakthrough) would be detecting oceans on half of the observed HZ rocky exoplanets. 


\subsubsection{Clouds} 
\label{sec:par1.3}

Refractive index and size of cloud particles allow for identification of their chemical composition and physical conditions in the planetary atmosphere, as has been demonstrated for Solar system planets. These parameters can be determined, for example, from the rainbow scattering angle, at which the rainbow polarization peaks in different wavelengths \citep[Fig.~\ref{fig:sc1},][]{Vaughan2023} or cloudbow polarization feature \citep[e.g.,][]{Emde2017, Sterzik2020, Roccetti2025a}. Retrieval of cloud cover and optical thickness is also feasible \citep[e.g.,][Roccetti et al. 2025c, submitted]{Roccetti2025a}.

Example scenarios to be tested for this parameter could be the same as for the atmospheres (Section~\ref{sec:par1.1}). However, PHPs with clear atmospheres, oceans and without clouds may also exist.

\subsection{Occurrence rate and type of PSLife (Survey 2)}
\label{sec:par2}

The existence of PSLife is currently known only for terrestrial life. The high linear polarization has been detected in various biopigment absorption bands within the ONIR range \citep{Berdyugina2016}. Accordingly, in Survey 2, the following surface components will be searched for on the PHPs identified in Surveys 1: exoBPs, ocean, land, ice (snow), and clouds. This will be achieved by measuring wavelength-dependent linear polarization during one planetary rotation at one illumination aspect (near orbital quadrature). 

For each surface component the following observable characteristics integrated over the illuminated part of the planet will be obtained: variations of the surface fraction (filling factor), wavelength-dependent geometrical albedo, linear polarization, and refractive index.

These observables are necessary to identify the presence, occurrence rate and nature of exoBPs among other possible surface components. Using the models described in Section~\ref{sec:obj}. An analysis of these parameters will inform us about possible PSLife on PHPs, i.e., possible LWs.

It is important to emphasize that polarization signatures of biopigments are robust with respect to other spectral and polarization effects. In particular, other surface features, like the occurrence of specific minerals, can also induce variations in the intensity and polarization spectrum. However, biopigments can create much stronger polarization signals and with a different angular dependence of the polarization \citep{Berdyugina2016}. Moreover, through planetary mapping (Follow-up 3, Section~\ref{sec:par3}) a more comprehensive imaging information on the distribution of biotic and abiotic matter across the planet can be obtained. In addition, residual stellar speckles vary in terms of location and appearance as a function of wavelength and can thus induce false spectral signals at the location of the planet under study. However, these effects are expected to be random, and hence follow-up observations should be able to distinguish real features from instrumental effects.

The resulting LW candidates with preliminary detections of exoBPs will be passed to the Follow up 3 and 4 observations and studies.

\subsubsection{The occurrence rate of exoBPs} 
\label{sec:par2.1}

This occurrence rate will be determined as a ratio of the number of exoBP detections 
to the number of observed (visited) PHPs.
The presence of exoBPs is revealed by strong variations of the albedo, from very low (down to 0.1) to very high (up to 0.9), caused by broad and deep absorption bands ($\sim$50--100 nm), which are accompanied by strong variations of linear polarization (up to 80\%) and steep edges (see Fig.~\ref{fig:sc2}).

Example scenarios to be tested for this parameter:

$\bullet$ Incremental Progress (Enhancing) would be that PSLife is not detected in any of the qualified single-visit observations, i.e., the exoBP occurrence rate is $<$1 (upper limit). This could be an indication for a smaller ($<$20\%) surface area of exoBPs, which is equivalent to a false-negative result. In this case, more and/or better quality data should be acquired. 

$\bullet$ Substantial Progress (Enabling) would be 
one detection out of 2--3 visits, i.e., PSLife is detected in at least 1/3 (less than 1/2) of the qualified single-visit observations, implying that the exoBPs surface area may be $>$20\%. This will allow to collect initial statistics on the occurrence rate of PSLife.

$\bullet$ Major Progress (Breakthrough) would be determining the exoBPs occurrence rate with a high statistical confidence. The maximum possible rate could be 2--3 detections out of 2--3 visits, i.e.,
PSLife is detected in all qualified single-visit observations, implying all PHPS have PSLife and the exoBP surface area is $>$20\%. In this scenario, verifying possible false-positives will be imperative. Alternatively, the exoPB occurrence rate could be in the range of 0.1--0.3, implying that PSLife (as we know it) would occur on one tenth to one third of PHPs. This does not exclude that other forms of life can be still present on other PHPs.

Verifying other reasons for non-detections (e.g., large cloud or ice covers) would be possible in the Follow-up 3 (Section~\ref{sec:par3}).

\subsubsection{Types of exoBPs} 
\label{sec:par2.2}

Types of globally dominant exoBPs will be revealed by specific central wavelengths, width and edges of the observed absorption and polarization, which suggest probable molecular bonds. These are known for the terrestrial biopigments observed in the ONIR range.

Example scenarios to be tested for this parameter:

$\bullet$ Incremental Progress (Enhancing) would be when possible exoBP detections were not statistically significant (upper limits), so that robust identification would not be possible.

$\bullet$ Substantial Progress (Enabling) would be when some of the detected exoBPs could be identified with those similar to Earth-like analogues. Additional work for characterization of unidentified exoBPs would be needed, or an analogous false-positive should be considered.

$\bullet$ Major Progress (Breakthrough) would be when the detected exoBPs could be identified with those similar to Earth-like analogues and other known or novel macromolecules.

\subsubsection{Relations with host stars} 
\label{sec:par2.3}

Correlations of the occurrence rates and diversity of exoBPs with the stellar flux at different wavelengths, i.e., spectral irradiance, and magnetic activity will be searched for detected exoBPs. This includes the presently known Earth--Sun case.

Example scenarios to be tested for this parameter:

$\bullet$ Incremental Progress (Enhancing) would be that no correlations could be found due a lack of exoBP identifications. 

$\bullet$ Substantial Progress (Enabling) would be that several detections and identifications were successful and allowed for revealing marginal trends. Additional detections and identifications would be needed.

$\bullet$ Major Progress (Breakthrough) could be two-fold. First, if all detections and identifications were for stars of similar classes, it could be either a preference for certain exoBPs on such stars, or an observational bias because of a favorable contrast ratio for this stellar class. Second, if stellar spectral classes were sufficiently different, clear relations (or absence of such) could be established.

\begin{table*}[ht!]
    \centering
    \caption[Physical Parameters]{Example scenarios to be tested in the four observing programs for the physical parameters which are needed to detect and characterize PSLife on exoplanets. Detailed descriptions are provided in Sections indicated in the last column.}
    \label{tab:par}
    \begin{tabular}{lccccl}
        \noalign{\smallskip}
        \hline
        \noalign{\smallskip}
        {Physical Parameter} & {State of} & {Incremental} & {Substantial} & {Major}&{} \\
        {} & {the Art} & {Progress} & {Progress} & {Progress}&{Section} \\
        {} & {} & {(Enhancing)} & {(Enabling)} & {(Breakthrough)}&{} \\
        \noalign{\smallskip}
        \hline
        \noalign{\smallskip}
        \multicolumn{6}{c}{{\bf Survey 1: Habitability likelihood of HZ rocky exoplanets}}\\
        \noalign{\smallskip}
        \hline
        Occurrence rate of PHPs with&&&&&\\
        \hspace{5mm}Atmospheres& 
        a few&
        1/2& 
        2/3& 
        9/10&
    \ref{sec:par1.1}\\
        \hspace{5mm}Liquid surfaces& 
        --&
        1/4& 
        1/3& 
        1/2&
    \ref{sec:par1.2}\\
         \hspace{5mm}Clouds& 
        --&
        1/2& 
        2/3& 
        9/10&
    \ref{sec:par1.3}\\
        \hline
        \noalign{\smallskip}
        \multicolumn{6}{c}{{\bf Survey 2: The occurrence rate and type of PSLife}}\\
        \noalign{\smallskip}
        \hline
        Occurrence rate of exoBPs& 
        Earth&
        $<1$& 
        $<0.5$& 
        0.1--0.3&
    \ref{sec:par2.1}\\
        Types of global exoBPs& 
        Earth&
        --& 
        Earth-like& 
        various&
    \ref{sec:par2.2}\\
        Relation with host stars& 
        Earth--Sun&
        --& 
        possible& 
        clear&
    \ref{sec:par2.3}\\
        \hline
        \noalign{\smallskip}
        \multicolumn{6}{c}{{\bf Follow-up 3: Abundance and diversity of PSLife}}\\
        \noalign{\smallskip}
        \hline
        Surface area of exoBPs& 
        19\% Earth&
        $\ge$50\%& 
        $\ge$20\%& 
        $\ge$(2--5)\%&
    \ref{sec:par3.1}\\
        Types of local exoBPs& 
        Earth&
        not clear& 
        one dominant& 
        various&
    \ref{sec:par3.2}\\
        Ocean / land surface areas& 
        71\% / 29\% Earth&
        $\ge$50\% / $\le$50\%& 
        $\ge$20\% / $\le$80\%& 
        $\ge$5\% / $\le$95\%&
    \ref{sec:par3.3}\\
        Types of abiotic surfaces& 
        Sol.Syst.&
        Earth-like& 
        Sol.Syst.-like& 
        other&
    \ref{sec:par3.4}\\
        Cloud properties& 
        Sol.Syst.&
        Earth-like& 
        Sol.Syst.-like& 
        other&
    \ref{sec:par3.5}\\
        Correlations of the components& 
        Earth&
        Earth-like& 
        Earth/non-terran& 
        non-terran&
    \ref{sec:par3.6}\\
        \hline
        \noalign{\smallskip}
        \multicolumn{6}{c}{{\bf Follow-up 4: Homochirality biosignature}}\\
        \noalign{\smallskip}
        \hline
        Asymmetric macromolecules& 
        Earth&
        0& 
        $<1$& 
        1&
    \ref{sec:par4.1}\\
        DOCP& 
        Earth&
        $>$0& 
        $>$0& 
        $>$10$^{-5}$&
    \ref{sec:par4.2}\\
        Central wavelength& 
        Earth&
        optical& 
        ONIR& 
        UVOIR&
    \ref{sec:par4.3}\\
        \hline
        \hline
    \end{tabular}
\end{table*}

\subsection{Abundance and diversity of PSlife (Follow-up 3)}
\label{sec:par3}

To determine an abundance and diversity of PSLife on the LW candidates identified in Survey 2, we will evaluate a fraction of the surface covered by PSLife and identify local exoPBs (i.e.,  within specific surface regions). This is achieved by obtaining spatially-resolved multi-color 2D maps of the exoplanetary surfaces from the HWO data (Section~\ref{sec:obs3}). The 2D maps of LWs obtained in this follow-up program are analogs of direct multi-spectral images of Earth from space (see Fig.~\ref{fig:sc3}, left).

The relevant physical parameters will be determined directly from such maps. The abundance of PSLife will be determined by the surface area of exoBPs relative to the total planetary and land surface areas. The diversity of the PSLife will be determined by the occurrence rate of various exoBPs on each LW (including Earth) relative to the total number of terrestrial and exoBPs.

In contrast to traditional exoplanetary spectral retrievals, there are no initial assumptions on the composition of the surface. The EPSI inversion method \citep{BerdyuginaKuhn2019} will reconstruct a multi-color 2D albedo map where each spatial resolution element has a low-resolution reflectance and polarization spectrum. This reconstructed spectrum will be then compared with possible known terrestrial samples and theoretical models to determine types of exoPBs and minerals which are dominant on the surfaces of alien LWs (Fig.~\ref{fig:sc3}, right). Because such a spectrum is integrated over a small fraction of the planetary surface (2--10\%), the number of possible constituents within this area is significantly smaller as compared with the entire planetary surface. Therefore, the reliability of identification of surface components is significantly higher than in spectral retrievals over the entire visible surface.

Examples of the biotic and abiotic physical parameters which will be determined in the Follow-up 3 program are listed in Table~\ref{tab:par} and briefly discussed below.

\subsubsection{Surface area with various exoBP} 
\label{sec:par3.1}

On Earth, the total surface area covered by various types of vegetation at its seasonal maximum is about 19\% of the total planet surface area,
as revealed, e.g., by the Moderate Resolution Imaging Spectroradiometer (MODIS) instrument on NASA's Terra satellite (see NOAA webpage).

Example scenarios to be tested for this parameter:

$\bullet$ Incremental Progress (Enhancing) would be resolving surface areas significantly larger than those on Earth (e.g., $\ge$50\%) with sparsely populated PSLife containing exoBPs. 

$\bullet$ Substantial Progress (Enabling) would be resolving surface areas comparable to those on Earth (e.g., $\ge$20\%) covered predominantly by PSLife containing exoBPs.

$\bullet$ Major Progress (Breakthrough) would be resolving compact (subcontinental) surface areas as those on Earth (e.g., $\ge$2--5\%) which are densely populated PSLife containing exoBPs.

\subsubsection{Types of dominant exoPBs for each specific area} 
\label{sec:par3.2}

On Earth, there known more than 200 photosynthetic and axillary biopigments.

Example scenarios to be tested for this parameter:

$\bullet$ Incremental Progress (Enhancing) would be a probable detection of one or more surface areas with exoBP but no clear identification.

$\bullet$ Substantial Progress (Enabling) would be detection of at least one surface area with a dominant exoBP, possibly similar to those known on Earth.

$\bullet$ Major Progress (Breakthrough) would be detection of multiple areas with dominant exoBPs, both Earth-like and non-terran.

\subsubsection{Ocean and land surface areas}
\label{sec:par3.3}

On modern Earth, ocean surface area is about 71\% of the total planetary area, while the rest 29\% is the land. Thus, the current terrestrial ocean-to-land area ratio is about 7/3, including areas permanently covered by ice and snow ($\sim$10\%, MODIS). This ratio has been evolving on Earth since its formation. Also the fraction of abiotic land has been varying in time. 

Example scenarios to be tested for this parameter:

$\bullet$ Incremental Progress (Enhancing) would be resolving ocean and land surface areas , e.g., $\ge$50\% and $\le$50\% for either.

$\bullet$ Substantial Progress (Enabling) would be detection of water (or other liquid) and land surface areas comparable with those on modern Earth, e.g., $\ge$20\%, similar to the land area on Earth.

$\bullet$ Major Progress (Breakthrough) would be detection of water (or other liquid) surface areas significantly smaller than those on modern Earth (e.g., $\ge$(2--5)\%, similar to subcontinental land area on Earth).

\subsubsection{Types of dominant abiotic constituents}
\label{sec:par3.4}

Abiotic surface composition is known for all planets, the majority of the moons and dwarf planets, and many asteroids in the Solar system. The known surface components include various liquids (e.g., Earth, Titan), minerals, organics and ices.

Example scenarios to be tested for this parameter:

$\bullet$ Incremental Progress (Enhancing) would be finding surface liquids, minerals, organics and ices, which are typical for Earth.

$\bullet$ Substantial Progress (Enabling) would be finding a variety of surface components typical for other objects in the Solar system. 

$\bullet$ Major Progress (Breakthrough) would be finding a large diversity of surface components, including those which are not typical for Earth and Solar system objects.

\subsubsection{Cloud fraction, patterns and particles}
\label{sec:par3.5}

Cloud cover fractions, patterns and particle types are known for all Solar system planets and moons which possess atmospheres. This includes various cloud particles properties, such as their size, refraction index and chemical composition. 

Example scenarios to be tested for this parameter:

$\bullet$ Incremental Progress (Enhancing) would be detecting clouds typical for Earth, e.g., with 10$\mu$m water droplets.

$\bullet$ Substantial Progress (Enabling) would be detecting water clouds with droplet sizes different from that on Earth.

$\bullet$ Major Progress (Breakthrough) would be detecting a large variety of cloud patters, droplets and aerosols, e.g., such as found in the atmospheres of Solar system objects, more massive exoplanets, and beyond.

\subsubsection{Correlations of exoBP abundance with other components}
\label{sec:par3.6}

Correlations of the PSLife type and abundance with the atmospheric and surface composition is known for Earth at different eons.

Example scenarios to be tested for this parameter:

$\bullet$ Incremental Progress (Enhancing) would be finding correlations typical for Earth.

$\bullet$ Substantial Progress (Enabling) would be finding correlations which are both typical and atypical for Earth.

$\bullet$ Major Progress (Breakthrough) would be finding only correlations which are atypical for Earth, e.g., possibly a different type of life origin.


\subsection{Homochirality biosignature (Follow-up 4)}
\label{sec:par4}

Circular polarization (CP) will be employed to search for a homochirality signature as a universal and unambiguous signature of life on Earth. Th search will be carried out for the best known to date LWs identified in the Follow-up 3 program.

All considerations for the Follow-up 4 program are made with the assumption that HWO will be serviceable and the coronagraph can be equipped with a circular polarimeter after identification of suitable targets through linear polarization measurements. Measuring CP signatures of exoplanets is unprecedented. Thus, observing even one LW of any size and around any star-type will be a substantial breakthrough. Since the degree of circular polarization (DOCP) will be very small, a trade-off between star luminosity and orbital distance should be evaluated carefully. 

The relevant physical parameters are the presence of asymmetric macromolecules, maximum DOCP, its central wavelength and other spectral characteristics of CP features. These are listed in Table\ref{tab:par} and briefly discussed below.

\subsubsection{Presence of asymmetric macromolecules}
\label{sec:par4.1}

Asymmetric macromolecules, biopigments and comparable molecules in CP have been detected only for terrestrial life. Small chiral organic molecules were also detected on meteorites originating in the Solar system.

Example scenarios to be tested for this parameter:

$\bullet$ Incremental Progress (Enhancing) would be a single target observations with a non-zero, narrow-banded CP signal (even with insufficient SNR of 2--3$\sigma$) which can indicate potential presence of life and justify further investigation and observation time. 

$\bullet$ Substantial Progress (Enabling) would be single target observations of a non-zero, narrow-banded CP signal with sufficient SNR of $\ge$5$\sigma$ which confirms the presence of PSLife (no known false-positives).

$\bullet$ Major Progress (Breakthrough) would be detecting more than one repeated observations and confirmations of a non-zero narrow-banded CP signal strengthen the finding of life on at least one LW.

\subsubsection{DOCP and handness}
\label{sec:par4.2}

An abiotic \textit{broad-banded} DOCP of 0--1\% have been detected for selected areas of Solar system planets, but not for exoplanets. A biotic \textit{narrow-banded} DOCP is known only for organisms on Earth with DOCP between 0.001--0.1\% (Fig.~\ref{fig:obs_sc4}.)

Example scenarios to be tested for this parameter:

$\bullet$ Incremental Progress (Enhancing) would be single target observation of a \textit{non-zero broad-banded} DOCP on exoplanets. The origin of this signature could be both abiotic and biotic.

$\bullet$ Substantial Progress (Enabling) would be single target observation of a \textit{non-zero narrow-banded} DOCP signal (even with insufficient SNR of 2--3$\sigma$) can indicate potential presence of homochiral biosignature.

$\bullet$ Major Progress (Breakthrough) would be detection of $>$0.001\% \textit{narrow-banded} DOCP which is expected for a planet with a major PSLife coverage. The CP sign would indicate the dominant handness in PSLife.

\subsubsection{Central wavelength of a narrow-banded CP feature}
\label{sec:par4.3}

CP signatures of terrestrial DNA/RNA, proteins and co-factors have been detected in UV--NUV and IR (in-situ), while CP of biopigments is observed in ONIR (remotely and in-situ).

Example scenarios to be tested for this parameter:

$\bullet$ Incremental Progress (Enhancing) would be if
central wavelengths of CP signatures were in the visible and similar to those of well characterized terrestrial biopigments. This could provide information on the nature of the exoPBs.

$\bullet$ Substantial Progress (Enabling) would be if central wavelengths of CP signatures were in the ONIR range and cover all known absorbance bands of organisms with biopigments on Earth, indicating a wider range of detectable organisms.

$\bullet$ Major Progress (Breakthrough) would be using UV spectropolarimetry which could reveal such macromolecules as DNA, RNA and proteins, which are universally abundant in terrestrial life.

\section{Description of Observations}
\label{sec:obs}



All science cases developed here aim at detecting reflected light from HZ rocky exoplanets in the ONIR using multi-wavelength (multi-band) imaging polarimetry combined with high-contrast coronagraphy. This justifies the need for HWP and required observational techniques and modes as discussed in this section for the proposed observing programs.

\begin{figure*}[ht!]
    \centering
    \includegraphics[width=1\textwidth]{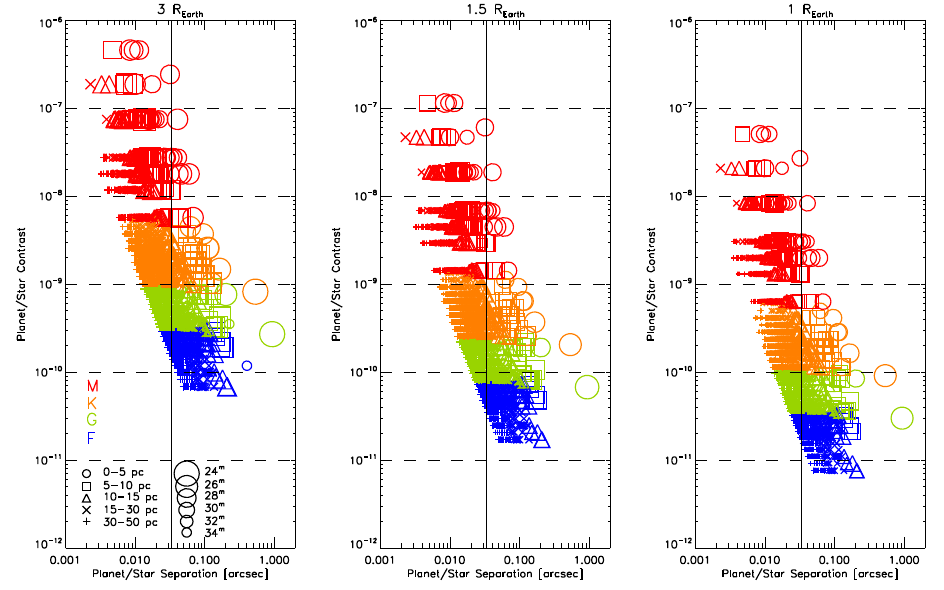}
    \caption{Planet/star contrast and angular separation for three hypothetical HZ rocky exoplanets of 1, 1.5, and 3 $R_{\rm Earth}$ which are potentially hosted by MKGF-type dwarf stars (potential host stars). These are shown within the 0--5, 5--10, 10--15, 15--30, and 30--50 pc distance bins and with the magnitude $V_{\rm star}\le13^{\rm m}$. For simplicity, the planet/star separation (the HZ distance) is scaled using the Earth/Sun separation. Symbols are colored according to the host star spectral class and sized with the exoplanet magnitude in reflected light, $24^{\rm m}\le V_{\rm refl}\le 34^{\rm m}$. The planet is assumed to be at an orbital quadrature ('half-moon' illumination phase) and to have an average geometrical albedo of 0.2 (similar to Earth). Different symbols correspond to different distance bins. The vertical solid line indicates the angular resolution of an 8m telescope (e.g., HWO) at the $2\lambda/D$ separation in $V$-band. Thus, the systems with the planet/star separations larger than this resolution limit (i.e., to the right from the vertical line) can be in principle resolved by such a telescope for each limiting planet/star contrast. Computed using the stellar data and approach from \citet{BerdyuginaKuhn2019}.}
    \label{fig:obs0}
\end{figure*}

\begin{table*}[ht!]
    \centering
    \caption[Observational Requirements]{Estimated number of extrasolar targets and detections (Earth is not included) with an 8m HWO telescope (see Fig.~\ref{fig:obs0}). The assumed occurrence rates are as follows: 0.3 for HZ rocky planets as an average for MKG dwarfs, 0.5 for PHPs among HZ rocky planets, and 0.2 for LWs among PHPs (see Table~\ref{tab:par} and discussion in Section~\ref{sec:par}). }
    \label{tab:obs0}
    \begin{tabular}{lcccc}
        \noalign{\smallskip}
        \hline
        \noalign{\smallskip}
        {Observation Requirement} & {State of} & {Incremental Progress} & {Substantial Progress} & {Major Progress} \\
        {} & {the Art} & {(Enhancing)} & {(Enabling)} & {(Breakthrough)} \\
        \noalign{\smallskip}
        \hline
        \noalign{\smallskip}
        Number of potential host stars& 
        &
        & 
        & 
        \\
        \hspace{5mm}with resolved 1.0 / 1.5 / 3.0 R$_{\rm Earth}$&
        & 
        14 / 40 / 59& 
        16 / 95 / 214& 
        67 / 136 / 519\\
        Host star spectral types& 
        --& 
        MK / MKG / MKGF& 
        MK / MKG / MKGF& 
        MK / MKG / MKGF\\
        Number of HZ rocky planets& 
        29 -- 70&
        4 / 12 / 18& 
        5 / 28 / 64& 
        20 / 41 / 156\\
        Number of detected PHPs (Survey 1)& 
        0& 
        2 / 6 / 9& 
        2 / 14 / 32& 
        10 / 20 / 78\\ 
        Number of detected LWs (Survey 2)&  
        0& 
        $<$1 / 1 / 2& 
        $<$1 / 3 / 6& 
        2 / 4 / 15\\
        Distance of LWs from the Sun [pc]& 
        --& 
        0--5& 
        0--15& 
        0--30\\
        Reflected light contrast at 2$\lambda/D$& 
        $\ge$10$^{-6}$& 
        $\le$10$^{-10}$& 
        $\le$10$^{-10}$& 
        $\le$10$^{-10}$\\ 
        \hline
        \hline
    \end{tabular}
\end{table*}

\subsubsection{Why HWO is needed?}
\label{sec:obs0.1}

HWO is the only large astronomical facility which is planned to achieve the yet unprecedented imaging contrast down to 0.1 ppb for exoplanets in stellar HZ within the solar neighborhood. This is at least four orders of magnitude better than 10$^{-6}$ that has been achieved with large ground-based telescopes. Therefore, only larger planets can be currently imaged in reflected light with such telescopes (Fig.~\ref{fig:obs0}).

As presented throughout this paper, multi-wavelength (multi-band) linear polarimetry combined with high-contrast imaging will enable robust discoveries of true PHPs and LWs, with surface oceans and PSLife, as well as first surface maps of life colonies on LWs. Circular polarimetry will verify whether homochirality is omnipresent in living organisms beyond Earth.

Preliminary estimates for the number of suitable targets which HWO could observe in ONIR reflected light within 50\,pc is presented in Fig.~\ref{fig:obs0} for three hypothetical HZ rocky exoplanets of 1, 1.5, and 3 $R_{\rm Earth}$ potentially hosted by MKGF-type dwarf stars (potential host stars) with the magnitude $V_{\rm star}\le13^{\rm m}$. For simplicity, the planet/star separation (the HZ distance) is scaled from the Earth/Sun separation. The planet is assumed to be observed at an orbital quadrature ('half-moon' illumination phase) and to have the average geometrical albedo 0.2 (similar to Earth). The angular resolution of an 8m HWO telescope capable of achieving the planet/star contrast of 10$^-{10}$ at the $2\lambda/D$ separation in $V$-band identifies initial targets. This is computed using the stellar data for 3526 stars in the Solar neighborhood (SIMBAD Database) and the approach employed by \citet{BerdyuginaKuhn2019}. The specific numbers are summarized in Table~\ref{tab:obs0}.
More detailed computations, including evaluation of exposure times for achieving the required signal-to-noise ration (SNR), are to be carried out with the HWO exposure time calculator for given telescope and coronagraph designs.

\subsubsection{Observational techniques}
\label{sec:obs0.2}

This section presents an overview of requirements for each observational technique required to implement the proposed Surveys and Follow-up programs.

{\bf Polarimetry.} 
Linear polarization mode (Stokes parameters I, q=Q/I, u=U/I) is needed for the science cases SC1--3. The required quality of polarimetric measurements is similar but not identical. This is presented for each case individually in Sections ~\ref{sec:obs1}--\ref{sec:obs3}.

Full Stokes polarization mode (Stokes parameters I, q=Q/I, u=U/I, v=V/I) is needed for SC4. This is a challenging observation, imposing additional requirements on the measurement of Stokes v (see Section~\ref{sec:obs4}).

Definitions of the polarimetric quality of measurements are as follows. A relative polarimetric sensitivity (precision), i.e., the photon noise level and noise-like speckle effects, depends on the flux of the planet at the achieved contrast. An absolute polarimetric accuracy is most likely limited by the accuracy of calibrating instrumental polarization effects that modify the polarimetric zero points. To reduce the bias due to a larger uncertainty in the polarimetric accuracy, it is essential that measurements at all wavelengths are taken simultaneously, under the same observing conditions. Calibrations on targets with a known polarization degree are to be carried out before and after the scientific measurements. 

{\bf Multi-band photometry or low-resoltuion spectroscopy.} 
Simultaneous ONIR observations (300--1000\,nm) with the resolution R$\sim$50 would be most optimal. Simultaneous ONIR multi-band measurements (with filter bands of 10\%) in the range of 300--1000\,nm would be acceptable. NUV spectropolarimetry from 200\,nm would be useful for a broader range of biomarkers (e.g., DNA, RNA, proteins) Alternatively, contemporaneous broad-band UV photometry would be useful for assessing biomarker retrieval.

{\bf Coronagraphy.} 
Direct images should provide a high enough contrast to directly image Earth-sized PHPs (Fig.~\ref{fig:obs0}):
the contrasts of 10$^{-6}$ to at least 10$^{-10}$ would allow for detections of PHPs hosted by main-sequence stars from M- to F-type, depending on the size and reflection properties of exoplanets, as well as the stellar magnitude and distance.

\subsubsection{Observing modes}
\label{sec:obs0.3}
The following observing modes are needed for the proposed Surveys and Follow-up programs.

{\bf Surveys of exoplanets.} 
For Survey 1, minimum 2--3 visits per target at preferred orbital phases are required to detect atmospheres, oceans and clouds in linear polarization and identify PHPs. Up to at least 30 known HZ rocky planets are desired for this survey to detect up to 10 PHPs. For Survey 2, minimum 1 visit per PHP near orbital quadrature is necessary to detect LWs with PSLife. Since organisms may occupy an unknown location and fraction of the visible exoplanetary surface, measurements should be carried out during one full axial planetary rotation. Up to at least 10 of promising PHPs from Survey 1 are desired for successful detection of at least one or more candidate LWs.

{\bf Follow-up time-series observations.} 
For Follow-up 3 program, minimum 10--20 visits at selected orbital phases per LW candidate are needed. Each visit consists of measurements during one full axial planetary rotation. A few most promising LW targets are desired from Surveys 1 and 2. Additional longer follow-up observations may be scheduled depending on the results of the initial studies to obtain a higher-resolution surface map of the most promising LW targets. For Follow-up 4 program, minimum 2--3 visits per LW target are needed to acquire full Stokes polarimetric measurements for at least one most promising LW identified in Follow-up 3. Measurements are to be carried out at the orbital and rotation phases where the linear polarization will peak due to the exoBP presence on the planetary surface.

\subsection{Detecting PHPs (Survey 1)}
\label{sec:obs1}

\subsubsection{Scattering angles}
\label{sec:obs1.1}

For each target, 2--3 visits are necessary to detect at least two critical characteristics of PHPs (Fig.~\ref{fig:sc1}):

{\bf Atmosphere:} first visit at a scattering angle between 90--120$^\circ$, i.e., near orbital quadrature when about a half to quarter of the planetary disk is illuminated by the star. At these angles, the continuum polarization due to Rayleigh scattering on atmospheric gases peaks in the blue, while in molecular bands it peaks at other wavelengths (e.g., H$_2$O and O$_2$ bands in green and red).

{\bf Ocean glint:} second visit at a scattering angle between 20--90$^\circ$, i.e., when only a small part of the planetary disk (crescent) is illuminated by the star. Here, the polarization due to ocean glint peaks in the red continuum, if ocean is cloud free.

{\bf Clouds:} third visit at a scattering angle between 130--150$^\circ$, i.e., when a rainbow polarization signature is expected. This is desirable but not critical for achieving the goal of Survey 1, namely to identify PHPs as candidate targets for Survey 2. This is because the cloud characterization is an immediate outcome of the Survey 2 and Follow-up 3 programs, where multiple observations of PHPs will be carried out at different scattering angles and axial rotation phases. The rainbow angles presented in Fig.~\ref{fig:sc1} are computed for water droplets in the Earth-like atmosphere. Should they be different in PHPs, this will be found out in the Follow-up 3 observing program.

The orbital phases for the specific scattering angles ($\phi$) mentioned above depend on the eccentricity of the orbit. For a circular orbit, the relation is 180$^\circ$--$\phi$. The Survey 1 observations can be carried out at any single planetary axial rotation phase, because of the assumed global presence of the atmosphere and ocean. If a significant cloud cover prevents clear detection of the ocean in Survey 1, observations of such PHPs should be repeated in Survey 2 at different axial rotation phases.

\subsubsection{Spectral coverage and resolution.}
\label{sec:obs1.2}

As shown in Fig.~\ref{fig:sc1}, in addition to specific scattering angles, the wavelength dependence of the polarization is crucial for recognizing signatures of Rayleigh scattering, ocean glint and rainbow. The ocean glint could be measured through the intersection of the phase curves at different broad-band channels across the ONIR range. For this purpose, the spectral resolution is not critical. However, the planet's rotation and changing weather patterns on the planet may complicate the interpretation of broad-band phase curves obtained at different times. Therefore, simultaneous measurements are desired in four spectral bands centered at 350, 450, 550 and 850 nm. However, even two simultaneous measurements at about 400 nm and 800 nm will be informative for the initial survey. A low-resolution spectropolarimetry (R$\sim$50) would be useful for determining gas composition of the atmosphere. 

\subsubsection{Polarimetric sensitivity}
\label{sec:obs1.3}

For the terrestrial atmosphere, the linear polarization degree for the PHP specific features is of the order of few tens \% (Fig.~\ref{fig:sc1}): 

$\bullet$ Rayleigh scattering: about 40\% at 350 nm,

$\bullet$ Ocean glint: about 35\% at 850 nm,

$\bullet$ Rainbow: about 20\% within the range of 350-850 nm.

Thus, the polarimetric precision (relative noise level) required for detecting such features is about 5\%. The polarimetric accuracy (absolute level of polarization) can be relaxed by a factor of 2--5 because of the specific wavelength dependence of the polarization degree, i.e., it can be in the range of 10--20\%, but polarimetric measurements at all wavelengths are to be taken simultaneously, under the same observing conditions.

The constraints listed above are summarized in Table~\ref{tab:obs}. They can be employed to determine the required wavelength-dependent photon rate and SNR. Information on possible targets is summarized in Table~\ref{tab:obs0}.

We note that a specific gas composition of the atmosphere has no effect on the constraints listed above (except for the clouds). However, a larger optical thickness of the atmosphere will reduce the atmosphere transparency and the peak Rayleigh polarization due to multiple scattering, especially in the blue. Therefore, achieving a higher polarimetric sensitivity would allow for identifying a broader range of PHPs -- also those with denser atmospheres than on Earth.


\begin{table*}[ht!]
    \centering
    \caption[Observational Requirements]{Requirements for the four observing programs to detect and characterize PSLife on exoplanets (see Section~\ref{sec:obs}).}
    \label{tab:obs}
    \begin{tabular}{lcccc}
        \noalign{\smallskip}
        \hline
        \noalign{\smallskip}
        {Observation} & {State of} & {Incremental Progress} & {Substantial Progress} & {Major Progress} \\
        {Requirement} & {the Art} & {(Enhancing)} & {(Enabling)} & {(Breakthrough)} \\
        \noalign{\smallskip}
        \hline
        \noalign{\smallskip}
        \multicolumn{5}{c}{{\bf Survey 1: Detecting PHPs}}\\
        \noalign{\smallskip}
        \hline
        Orbital scat. angles: & 
        only in&
        & 
        & 
        \\
        \hspace{5mm}atmosphere =& 
        the Solar& 
        = 90--120$^\circ$& 
        = 90--120$^\circ$& 
        = 90--120$^\circ$\\
        \hspace{5mm}ocean glint =& 
        system& 
        & 
        = 20--40$^\circ$& 
        = 20--40$^\circ$\\
        \hspace{5mm}rainbow =& 
        & 
        & 
        & 
        = 130--150$^\circ$\\
        Rotational phases =& 
        = any& 
        = any& 
        = any& 
        = any\\
        \hline
        Spectroscopy:& 
        Neptune& 
        400, 800 nm& 
        350,450,850 nm& 
        350,450,550,850 nm\\
        multi-band& 
        \& Jupiter& 
        20\% width& 
        20\% width& 
        10--15\% width or\\
        simultaneously& 
        size exopl.& 
        & 
        & 
        R$\sim$50 resolution\\
        \hline
        {\em Linear} Polarimetry: & 
        Few ppm&
        & 
        & 
        \\
        \hspace{5mm}precision =& 
        for hot& 
        = 5\%& 
        = 5\%& 
        = 2\%\\
        \hspace{5mm}accuracy =& 
        Jupiters& 
        = 20\%& 
        = 10\%& 
        = 5\%\\
        \hline
        \noalign{\smallskip}
        \multicolumn{5}{c}{{\bf Survey 2: Detecting LWs}}\\
        \noalign{\smallskip}
        \hline
        Orbital scat. angles: & 
        only on&
        = 90--120$^\circ$& 
        = 90--120$^\circ$& 
        = 90--120$^\circ$\\
        Rotational phases =& 
        Earth& 
        = 0--360$^\circ$& 
        = 0--360$^\circ$& 
        = 0--360$^\circ$\\
        \hline
        Spectroscopy:& 
        only on& 
        550 nm& 
        450,550,850 nm& 
        450,550,650,850 nm\\
        multi-band& 
        Earth& 
        20\% width& 
        20\% width& 
        10--15\% width or\\
        simultaneously& 
        & 
        & 
        & 
        R$\sim$50 resolution\\
        \hline
        {\em Linear} Polarimetry: & 
        0.1--0.01\%&
        & 
        & 
        \\
        \hspace{5mm}precision =& 
        on Earth& 
        = 5\%& 
        = 2\%& 
        = 1\%\\
        \hspace{5mm}accuracy =& 
        & 
        = 5\%& 
        = 5\%& 
        = 2\%\\
        \hline
        \noalign{\smallskip}
        \multicolumn{5}{c}{{\bf Follow-up 3: Characterizing LWs}}\\
        \noalign{\smallskip}
        \hline
        Orbital scat. angles: & 
        only on&
        = 60--180$^\circ$& 
        = 45--315$^\circ$& 
        = 0--360$^\circ$\\
        Rotational phases =& 
        Earth& 
        = 0--360$^\circ$& 
        = 0--360$^\circ$& 
        = 0--360$^\circ$\\
        \hline
        Spectroscopy:& 
        only& 
        550 nm& 
        450,550,850 nm& 
        450,550,650,850 nm\\
        multi-band& 
        on Earth& 
        20\% width& 
        20\% width& 
        10--15\% width or\\
        simultaneously& 
        & 
        & 
        & 
        R$\sim$50 resolution\\
        \hline
        {\em Linear} Polarimetry: & 
        0.1--0.01\%&
        & 
        & 
        \\
        \hspace{5mm}precision =& 
        on Earth& 
        = 5\%& 
        = 2\%& 
        = 1\%\\
        \hspace{5mm}accuracy =& 
        & 
        = 5\%& 
        = 5\%& 
        = 2\%\\
        \hline
        \noalign{\smallskip}
        \multicolumn{5}{c}{{\bf Follow-up 4: Verifying homochirality on LWs}}\\
        \noalign{\smallskip}
        \hline
        Orbital scat. angle & 
        only on&
        \multicolumn{3}{l}{at the phase where the maximum exoPB abundance}\\
        \& rotational phase& 
        Earth& 
        \multicolumn{3}{l}{ is detected in Follow-up 3}\\
        \hline
        Spectroscopy:& 
        only on& 
        700 nm& 
        600--750 nm& 
        600--750 nm\\
        & 
        Earth& 
        10\% width& 
        R$\sim$50 & 
        R$\sim$100 \\
        \hline
        {\em Circular} Polarimetry: & 
        $\sim$0.01\%&
        & 
        & 
        \\
        \hspace{5mm}precision =& 
        on Earth& 
        = 0.1\%& 
        = 0.03\%& 
        = 0.01\%\\
        \hspace{5mm}accuracy =& 
        & 
        = 1\%& 
        = 0.5\%& 
        = 0.1\%\\
        \hline
        \hline
    \end{tabular}
\end{table*}

\begin{figure*}[ht!]
    \centering
    \includegraphics[width=0.75\textwidth]{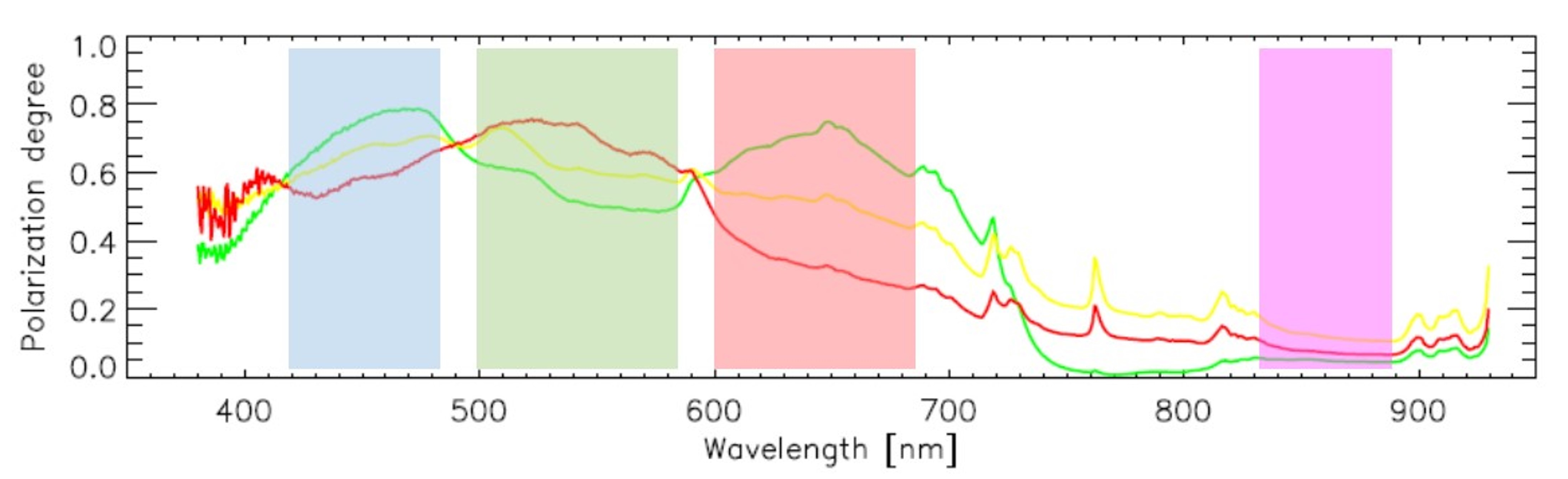}
    \caption{
Linear polarization degree detected in reflected light from oxygenic photosynthetic organisms (here, plants) for pure samples with dominant biopigments: chlorophyll (green), carotenoids (yellow), anthocyanins (red). The high polarization degree up to 80\% is dominant in the optical 400--700 nm spectral range, while it is reduced down to a few \%  in the NIR. The wavelength regions preferred for broad-band flux polarization measurements (or low-resolution spectropolarimetry) are marked as solid-color bands (blue, green, orange, magenta).  From \citet{Berdyugina2018}.
}
    \label{fig:obs_sc2}
\end{figure*}

\begin{figure*}[ht!]
    \centering
    \includegraphics[width=0.7\textwidth]{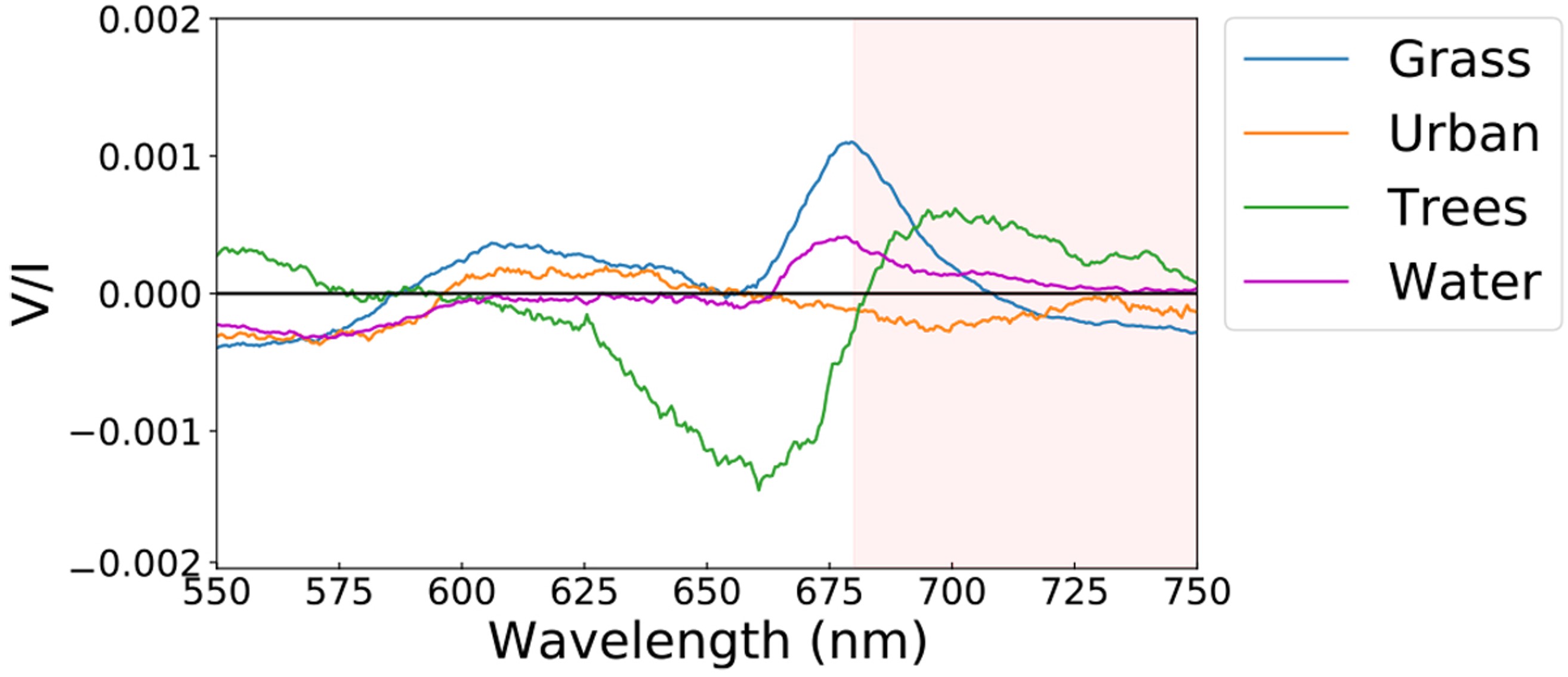}
    \caption{
Circular polarization spectra for various landscape features: grass, urban area, trees, and water containing algae obtained with a polarimeter on a helicopter. The red shaded area represents the red edge. From \citet{Patty2021}.
    }
    \label{fig:obs_sc4}
\end{figure*}

\subsection{Detecting Living Worlds (survey 2)}
\label{sec:obs2}

For the PHPs with a high habitability likelihood identified in Survey 1, we propose to carry out a second survey covering one full axial rotation for each target. 

The following observables are needed for detecting LWs with exoPB signatures and identifying their nature: absorption depth and polarization amplitude (\%), central wavelength (nm), full width at half maximum (FWHM) of the absorption and polarization bands (nm), steepness of the biopigment “edges”: maximum gradients (\%/nm).

{\bf Time series.}
For each PHP target, one visit near an orbital quadrature (partial illumination) is necessary for obtaining the above observables at 5--10 evenly distributed planetary axial rotation phases. This is necessary for detecting biogenic and non-biogenic surface components (see Section~\ref{sec:par3.2}).

{\bf Spectral coverage and resolution.}
The following wavelength regions are chosen to identify exoPB absorption bands similar to those occurring in the ONIR region in terrestrial 
photosynthetic organisms (Fig.~\ref{fig:obs_sc2}).

    $\bullet$ High-contrast coronagraphic broad-band linearly-polarized flux measurements in at least four bands (quasi-)simultaneously: 420--480 nm, 500--580 nm, 600--690 nm, 830--890 nm.
    
    $\bullet$ Alternatively, high-contrast coronagraphic low resolution linear spectropolarimetry (R$\sim$50--100), (quasi-)simultaneously in at least three bands: 400--550 nm, 550--700 nm, 700--850 nm.

If none of the known biopigments can be identified, the presence of deep, broad, highly-polarized bands will indicate a possible presence of unknown exoPBs. Multi-wavelength (spectro-)polarimetry is the only remote-sensing tool capable of clarifying their nature. 

{\bf Polarimetric sensitivity.}
The linear polarization degree of pure terrestrial samples is very high, up to 80\% (see Fig.~\ref{fig:obs_sc2}). However, it is expected that exoPBs cover only a fraction of the exoplanetary surface which is visible at a given orbital-rotational (illumination) phase. Therefore, polarimetric precision of 2--5\% (Table 4.0) is required in Stokes q and u to detect 10-20\% surface coverage by biopigments.


\subsection{Characterizing Living Worlds (Follow-up 3)}
\label{sec:obs3}

For the most promising LW candidates with preliminary detections identified in Survey 2, we propose to carry out follow up observations to obtain (i) multi-color surface maps of LWs and (ii) surface-resolved (spatially-resolved) absorption and polarization spectra to identify various surface features described in Section~\ref{sec:par3.3}. 

Observables, spectral and polarimetric requirements are the same as for Survey 2 (Section~\ref{sec:obs2}), but data acquisition time series are different. 

{\bf Time series.} 
For each LW candidate, at least 10--20 visits are required at orbital phases (scattering angles) where the presence of exoBPs was identified as most conspicuous in Survey 2. Each visit consists of 5--10 evenly distributed planetary axial rotation phases. This is necessary to increase the spatial resolution of the surface maps \citep{BerdyuginaKuhn2019}. 

Examples of spatially resolved spectra measured in four ONIR passbands which could be inferred from the map are shown in Fig.~\ref{fig:sc3}. Such spectra clearly distinguish between biogenic and non-biogenic components (see Section~\ref{sec:par3}), when spatial resolution elements are relatively homogeneous in their composition. For detecting continents and oceans and determining the ocean-to-land ratio on PHPs without exoBPs, a smaller number of observations (about 50) would be sufficient, also for tidally-locked exoplanets.


\subsection{Verifying homochirality on LWs (Follow-up 4)}
\label{sec:obs4}

For the LWs where PSLife and exoBPs were detected with a high confidence in spatially resolved surface maps (Follow-up 3), we propose to carry out additional follow up observations to evaluate the presence of homochirality in the detected alien organisms. This is a challenging observational program in terms of the required quality of polarization measurements, which may only be possible with an upgraded HWO coronagraph equipped with a high-precision full-Stokes polarimeter. Nevertheless, for the sake of completeness, we present here the observational requirements for this program as well. 

{\bf Scattering angles.}
For each target, probably at least 2--3 visits are needed for the same orbital and axial rotation phase to collect enough photons for the required polarimetric sensitivity. The most optimal scattering angle and rotational phase will be selected using the LW maps from the Follow-up 3 program, in particular, where the maximum exoBP abundance was detected.

{\bf Spectral coverage and resolution.}
Measurements shown in Fig.~\ref{fig:obs_sc4} demonstrate that the width of the spectral bands should be 20--50 nm, or alternatively the spectral resolution should be minimum 30, but preferably 100.

{\bf Polarimetric sensitivity.}
For the light reflected off dense areas of photosynthetic organisms, the degree of circular polarization is generally of the order of 0.1--0.01\%, but can reach up to 1\%, independent from the phase angle of incident light in microbial mats \citep{Sparks2021}, in algae \citep{Patty2018}, and directional \citep{Patty2022}. This is confirmed by measurements made in both laboratory experiments and remote sensing operations from a helicopter and hot-air balloon \citep{Patty2021, Mulder2022} (see examples in Fig.~\ref{fig:obs_sc4}. Therefore, obtaining such measurements for exoplanets requires the following:

    $\bullet$ a relative polarimetric sensitivity (precision) for CP (Stokes v) of better than 0.1--0.03\% (noise level);
    
    $\bullet$ an absolute polarimetric accuracy for Stokes v of better than $\sim$1–0.1\%, to provide a baseline for spectral variations, including zero-polarization crossings of v at given wavelengths;

    $\bullet$ calibration accuracy for cross-talk (q,u)$\rightarrow$v of better than $\sim$1\%, to ensure that the much stronger linear polarization does not induce false circular polarization signals.

\section{Conclusions}
\label{sec:con}

Multi-wavelength polarimetry of HZ rocky exoplanets imaged at an unprecedented contrast using the HWO coronagraph is a novel opportunity for a robust discovery of life on exoplanets.
The observing programs proposed here aim at detecting and characterizing alien Living Worlds with photosynthetic life and determining their occurrence rate.

Initial identification of potentially habitable planets through detections of their atmospheres, clouds and liquid surface water (ocean) will provide targets for further identification of Living Worlds and obtaining their multi-color surface maps using linear polarimetry. 
The surface distribution, abundance and types of alien photosynthetic organisms with exo-biopigments will be determined from these maps. Eventually, chiral properties of alien biomolecules could be determined using circular polarimetry.

Ultimately, a quantitative answer to the question: "Are we are alone in the Universe?" will be obtained.


\bibliography{hwo_polarimetry.bib}

\end{document}